\documentclass{article}
\usepackage{newtxtext}
\usepackage{newtxmath}
\usepackage{graphicx}
\usepackage{xcolor}
\usepackage{booktabs}
\usepackage{subcaption}
\usepackage{multirow}
\usepackage{siunitx}
\usepackage{bm}

\usepackage[english]{babel}

\usepackage[letterpaper,top=2cm,bottom=2cm,left=3cm,right=3cm,marginparwidth=1.75cm]{geometry}

\usepackage[colorlinks=true, allcolors=blue]{hyperref}

\title{Learning Filters in Feedback Delay Networks from Noisy Room Impulse Responses}
\author{Gloria Dal Santo\textsuperscript{1}\thanks{To whom correspondence should be addressed, e-mail: gloria.dalsanto@aalto.fi} \and Karolina Prawda\textsuperscript{2} \and Sebastian J. Schlecht\textsuperscript{3} \and Vesa V\"alim\"aki\textsuperscript{1}}

\newcommand{\mat}[1]{\bm{#1}}
\newcommand{\fs}[0]{f_\textrm{s}} 
\newcommand{\Tsixty}[0]{T_{60}}

\begin{document}
\maketitle
\begin{center}
	\textsuperscript{1}Acoustics Lab, Department of Information and Communications Engineering, Aalto University, Espoo, Finland\\
	\textsuperscript{2}AudioLab, University of York, York, United Kingdom\\
	\textsuperscript{3}Friedrich-Alexander-Universit\"{a}t Erlangen-N\"{u}rnberg (FAU), Erlangen, Germany
\end{center}
\abstract{%
Recursion is a fundamental concept in the design of filters and audio systems. In particular, artificial reverberation systems that use delay networks depend on recursive paths to control both echo density and the decay rate of modal components. The differentiable digital signal processing framework has shown promise in automatically tuning recursive and non-recursive elements using gradient-based optimization with perceptually or physically motivated loss functions, such as energy decay or spectrogram differences. These representations are highly sensitive to model mismatches, which can lead to spurious loss minima. In particular, discrepancies in background noise can result in inaccurate attenuation estimates. This paper addresses the problem of tuning recursive attenuation filters of a feedback delay network when targets are noisy. We analyze the loss profile associated with different optimization objectives and propose a method that explicitly models noise, improving the accuracy of the estimated attenuation filters under low signal-to-noise conditions. We demonstrate the effectiveness of the proposed approach through statistical analysis on both synthetic and real target data. Furthermore, we identify the sensitivity of attenuation filter parameters tuning to perturbations in frequency-independent parameters. These findings provide practical guidelines for more robust and reproducible gradient-based optimization of feedback delay networks.}
\maketitle
\section{INTRODUCTION}

Differentiable digital signal processing (DDSP)~\cite{engel2020ddsp, hayes2024review} for filter design is an approach that applies automatic differentiation to digital signal processing operations. By implementing filters in a differentiable manner, it becomes possible to backpropagate loss gradients through their parameters, enabling data-driven optimization without losing domain knowledge. 
DDSP has gained popularity due to these domain-appropriate inductive biases, yet it still presents several challenges~\cite{hayes2024review}. The parameters of differentiable filters are often constrained by stability conditions~\cite{kuznetsov2020differentiable}, may be subject to the responsibility problem~\cite{hayes2025audio}, and may span different domains and distributions, making optimization nontrivial. 

In artificial reverberation, the feedback delay network (FDN) became a particularly popular choice for late reverberation modeling due to its versatility and computational efficiency~\cite{jot1991digital, valimaki2012fifty}. More recently, the use of FDN has also gained traction within DDSP frameworks for blind reverb estimation~\cite{lee2022differentiable, gotz2025matching} and effects matching~\cite{yu2025diffvox}. The application of differentiable FDNs is also supported by research focused on gradient-based optimization without the use of neural networks. Instead of training a model, the algorithm directly fine-tunes the FDN's parameters to improve its perceptual quality~\cite{dal2023differentiable, santo2024optimizing, dal2024rir2fdn} or match the characteristics of a target room impulse response (RIR)~\cite{mezza2024data, mezza2024modeling, giampiccolo2025modeling, giampiccolo2024differentiable}. 

The design of an FDN typically follows a two-step process. First, a lossless FDN~\cite{rocchesso2002circulant} is constructed such that the input energy circulates through the feedback loop without attenuation, resulting in no decay over time. In the second step, attenuation filters are added to the feedback paths to control the rate of energy decay, thereby determining the system's reverberation time ($T_\textrm{60}$). While the latter is more crucial when matching a target RIR, given the perceptual role of $T_{60}$, poor lossless FDN designs can lead to coloration and sparseness issues that degrade the plausibility of the system's output. These issues typically manifest as metallic ringing and temporal roughness, respectively, causing the output to deviate from an ideal smooth late reverberation by introducing prominent resonances and temporal fluctuations. Our previous work~\cite{dal2023differentiable, santo2024optimizing} addressed this by optimizing the frequency-independent parameters using gradient descent to produce a perceptually colorless and temporally dense response. 

The design of the attenuation filter depends on several factors, including the desired control flexibility, computational complexity constraints, and required accuracy. One common requirement is to ensure that the filter strength, namely the per-sample gain, is proportional to the corresponding delay length, thereby ensuring a correct decay rate across modes. Early implementations primarily relied on low-order lowpass filters~\cite{moorer1979reverberation, jot1991digital, jot1992analysis, jot1997efficient}, which provided only limited control over a few frequency bands. Higher-order filters were later used to simulate the $T_{60}$ in octave bands, among which proportional graphic equalizers (GEQ) became widely popular due to their simplicity~\cite{jot2015proportional}, and availability of accurate designs~\cite{schlecht2017accurate, prawda2019improved, valimaki2024two}. 
These filter designs, however, assume that the target $T_{60}$ values are known. 

While standard methods exist for estimating $T_{60}$ directly from the RIR, typically based on Schroeder backward integration via the energy decay curve (EDC)~\cite{schroeder1965new}, these approaches often yield incorrect estimates. Such errors generally arise from mismatches with the assumed single-decay slope model, the presence of high background noise, or non-stationary noise~\cite{morgan1997parametric, karjalainen2002estimation, xiang2001evaluation, gotz2022neural}. It is therefore highly desirable to estimate the filter parameters directly from the measured RIR, or from the reverberated material itself. The latter poses a significantly more challenging problem that can benefit from recent advances in techniques such as DDSP and other deep learning-based methods. 

In this context, Lee et al.~\cite{lee2022differentiable} presented a parameter estimation network to learn a subset of FDN parameters, including those of a shared differentiable parametric equalizer (PEQ) implemented with state-variable filters. Ibnyahya et al. introduced a scalable, proportional, and differentiable PEQ that maintains accuracy and stability comparable to state-of-the-art informed methods~\cite{Ibnyahya2025}, and suggested its applicability to differentiable FDNs. Mezza et al.~\cite{mezza2024modeling} implemented an attenuation filter bank as a single one-dimensional convolutional layer, assigning a dedicated kernel to each delay-line output. The problem of jointly optimizing the attenuation filters and minimizing coloration and sparsity issues has been partially addressed by Lee et al.~\cite{lee2022differentiable}, who incorporated additional components, including learnable all-pass filters.

Recent work has demonstrated promising results by incorporating a loss term that regularizes the feedback matrix in FDNs. Specifically, G\"otz et al.~\cite{gotz2025matching} applied it in blind estimation using acoustic embeddings, while Das et al.~\cite{das2025differentiable} explored grouped FDNs for modeling coupled rooms. Both methods jointly optimized frequency-independent and frequency-dependent parameters through composite loss functions. Specifically, to optimize the attenuation filters,~\cite{gotz2025matching} uses a multiscale spectral (MSS) loss, while~\cite{das2025differentiable} includes a distance metric on EDCs. 
As noted by the authors in~\cite{dal2024similarity}, common time–frequency representations such as the MSS, although effective for many synthesis tasks, do not fully capture the statistical properties characteristic of RIRs. EDC-based losses, while they are more perceptually valid~\cite{dal2024similarity}, present asymmetries around the minimum and are highly sensitive to background noise and spectral discrepancies. Moreover, although attenuation is fully determined by the strength of the attenuating elements in the feedback loop, MSS and EDC representations are also influenced by variations in the frequency-independent elements, whose effects have not yet been examined.

Measured RIRs naturally contain background and measurement noise. Such noise can be reduced either by RIR truncation, where the noise-only portion of the signal is discarded \cite{karjalainen2002estimation, Lundeby_1995_uncertainties, abel2010methods, prawda2025_cropping}, or by averaging or median filtering \cite{prawda2024non}, which rely on multiple recordings of the same space. However, in the context of blind reverberation estimation based on DDSP, noise estimation and suppression, or explicit noise modeling, have received comparatively little attention. In \cite{steinmetz2021filtered}, RIRs are represented using a trained filterbank and temporal masks that are applied to shape a noise sequence, whereas in \cite{lee2022differentiable, gotz2025matching}, the impact of measurement noise is not addressed, and the learnable system does not model its presence. Although publicly available RIR datasets generally exhibit a high signal-to-noise ratio (SNR), mixed-reality applications, particularly those using microphones embedded in head-worn devices, often involve higher levels of measurement noise and background noise that can impair the discoverability of energy decay rates. Existing research on this topic has primarily focused on enhancing spatial properties rather than addressing the noise robustness of energy decay estimators \cite{van2008effect, fernandez2022enhancing}.

This paper investigates the performance of loss-function configurations for attenuation filters in FDNs.  We explicitly focus on loss functions based on the EDC and MSS representations of the RIR and quantify how their accuracy varies across two noise levels. \noindent The main contributions of this paper are summarized as follows:
 \begin{enumerate}
\item We present a systematic method to evaluate the accuracy and robustness of the loss functions in recovering ground-truth parameters.
\item We quantify the sensitivity of EDC- and MSS-based loss functions to background noise and frequency-independent variable changes.
\item We show that matching the target RIR SNR during optimization by injecting synthetic noise into the estimated response significantly reduces the mean absolute error between target and predicted energy decays.
\end{enumerate}
To make the loss profile interpretable, our study is restricted to a low-order filter design parameterized by only two variables: $\Tsixty$ at dc and the crossover frequency. Although methods for projecting higher-dimensional loss functions onto a 2D plane exist (e.g., \cite{li2018visualizing}), we demonstrate that valuable insights can be derived from this simpler configuration. The investigation of higher-order filters is reserved for future work.

The paper is structured as follows. Sec.~\ref{sec:background} provides background information on the FDN, and presents common loss functions used to learn its parameters. Sec.~\ref{sec:method} describes the system configuration and methodology used to analyze the effects of background noise and parameter perturbations on the loss profile and on performance under gradient-descent optimization. The details of these tests and their results are presented in Sec.~\ref{sec:evaluation}. The results are further discussed in Sec.~\ref{sec:discussion}. Finally, Sec.~\ref{sec:conclusion} offers concluding remarks. 

\section{BACKGROUND}\label{sec:background}

In this section, we present the FDN in its standard configuration, depicted in Fig.~\ref{fig:FDN_diagram}, which uses proportional filters in the recursive path to model the desired energy decay. In our study, we adopt the frequency-sampling method to implement a differentiable FDN and its attenuation filters, and we explain the main principles of this approach here. We then discuss a limitation inherent to the FDN response, namely its coloration and sparsity, and emphasize that mitigating these issues should not conflict with the optimization of the attenuation filters. We conclude this section by outlining a straightforward approach to visualizing the loss profile and by describing the loss functions under investigation.
\subsection{Feedback Delay Network}
\begin{figure}[t!]
\centerline{\includegraphics[trim={0.9cm 0.2cm 0.9cm 0.7cm}, clip, width=0.75\textwidth]{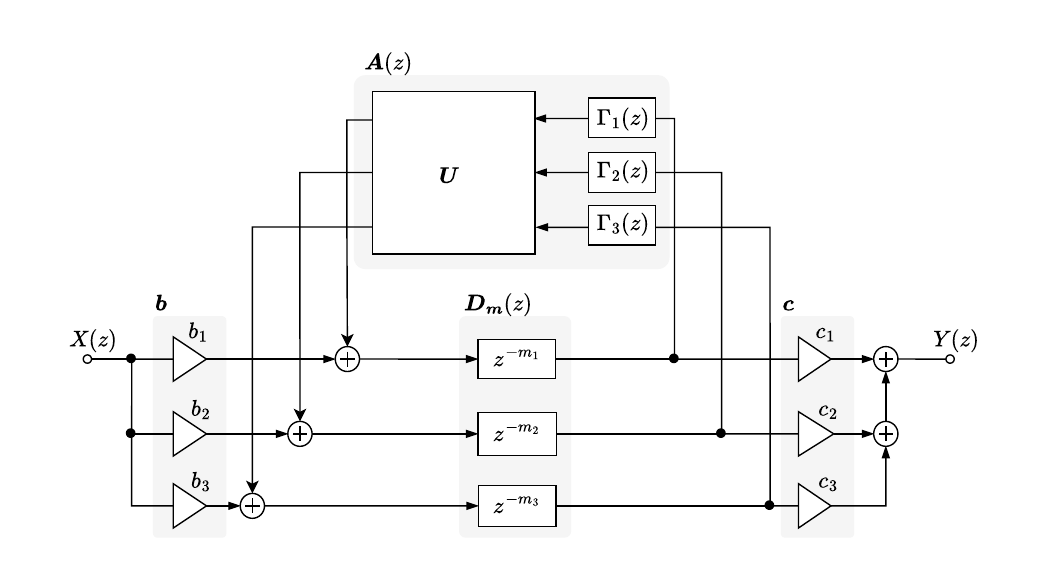}}
\caption{\textit{Structure of a SISO FDN with $N=3$ delay lines, where the attenuation filters are located in blocks $\Gamma_i(z)$.}}
\label{fig:FDN_diagram}
\end{figure}

An FDN is a recursive system consisting of delay lines, a set of gains, and a feedback matrix through which the delay-line outputs are coupled with the delay-line inputs, as shown in Fig.~\ref{fig:FDN_diagram}.
The transfer function of a single-input, single-output (SISO) FDN can be written as 
\begin{align}\label{eq:tr_fdn}
    H_\text{FDN}(z) = \frac{Y(z)}{X(z)} = \mat{c}^\top\big[\mat{D_m}(z)^{-1} -\mat{A}(z)\big]^{-1}\mat{b}\,, 
\end{align}
\noindent where $\mat{A}(z)$ is a filter feedback matrix consisting of an $N \times N$ mixing matrix and $N$ parallel filtering stages, where $N$ is the number of delay lines, vectors $\mat{b}$ and $\mat{c}$ are $N\times1$ column vectors of input and output gains, respectively, and the operator $(\cdot)^\top$ denotes the transpose. 
The vector $\mat{m} = [m_1, \dots, m_N]$ defines the lengths of the delay lines in samples and the system's order $\mathcal{{M}} = \sum_{i = 1}^Nm_i$. Its corresponding $N\times N$ delay matrix $\mat{D_m}(z)$ is created by taking a diagonal matrix with entries given by $[z^{-m_1}, \dots, z^{-m_N}]$.
The feedback matrix is defined as follows:
\begin{align}\label{eq:FFM}
    \mat{A}(z) = \mat{U}\mat{\Gamma}(z),
\end{align}
where $\mat{U}$ is an $N \times N$ orthogonal matrix and $\mat{\Gamma}(z)$ is a diagonal attenuation matrix, whose diagonal entries are the delay-line specific attenuation filters $\Gamma_i(z)$, with $i$ indexing the $i$-th delay-line.

\subsection{Modal Interpretation of the FDN Parameters}\label{subsec:param_role}
To better understand the contribution of each parameter in Eq.~\eqref{eq:tr_fdn}, Schlecht et al. examined the FDN through its modal decomposition~\cite{schlecht2019modal}. For orthogonal $\mat{U}$, the decay of the modes is governed entirely by the attenuation filters $\mat{\Gamma}(z)$. The modal frequencies are determined by both $\mat{U}$ and $\mat{D_m}(z)$, and statistical analyses have shown that, for randomly generated FDNs, these frequencies are nearly equidistributed~\cite{schlecht2019modal}. Although the initial phase of the modes is generally regarded as less perceptually relevant, the distribution of their initial magnitudes, referred to as modal excitation, has been shown to correlate with perceived coloration~\cite{heldmann2021role}. The modal excitation is influenced by all FDN parameters except the attenuation filters $\mat{\Gamma}(z)$.
These observations imply that adjusting the attenuating elements is sufficient to control the modal decay, but not sufficient to control spectral energy. Yet, the available modal decomposition algorithms would likely lead to excessive memory requirements and prohibitively long optimization times~\cite{schlecht2019modal}, limiting their use within iterative, loss-based optimization. Instead, we can optimize modal decay using the signal’s energy decay, which depends on both modal excitation and frequency distribution.

\subsection{Proportional Filter Design}
To ensure that all system modes decay at the same rate, corresponding to a target frequency-dependent reverberation time $\Tsixty(\omega)$, the strength of the attenuation filters must be proportional to the corresponding delay lengths, that is, longer delays require stronger attenuation~\cite{jot1991digital}. This proportionality allows the definition of a global per-sample attenuation that is valid regardless of the signal's path through the feedback loop. 

More specifically, for a target $\Tsixty (\omega)$ the prototype magnitude response of the attenuation filter, evaluated at $z=e^{\imath\omega}$, is 
\begin{align}
    \left| {\Gamma}_{\textrm{prot}}(e^{\imath\omega})\right|  =  10^{\frac{\gamma_{\textrm{dB}(\omega)}}{20}}  =  10^{-\frac{3}{ f_\textrm{s} \Tsixty(\omega)}}, 
\end{align}

\noindent where $\gamma_{\textrm{dB}}(\omega) = -60/f_\textrm{s} \Tsixty(\omega)$ denotes the per-sample attenuation in decibel (dB) and $f_\textrm{s}$ is the sampling rate.
The prototype response is then scaled according to the delay-line length $m_i$, resulting in a target attenuation filter for each delay line:
\begin{align}\label{eq:scaled_filters}
    {\Gamma}_i(e^{\imath\omega}) = \bigg|10^{-\frac{\gamma_{\textrm{dB}}(\omega)}{20 }}\bigg|^{m_i} = |{\Gamma}_{\textrm{prot}}(e^{\imath\omega})|^{m_i}.
\end{align}
This construction requires the filters to be \textit{proportional}, i.e., filters $\Gamma_i(z)$ must have a magnitude response in dB that can be obtained by scaling the prototype response by a frequency-independent factor. For self-similar filters \cite{abel2004filter}, which satisfy this proportionality condition, each attenuation filter ${\Gamma}_i(z)$ can be derived from the filter associated with the shortest delay line
, $m_\textrm{min}$, which is the most challenging filter to design in stable systems due to its magnitude response's proximity to 0 dB. Given ${\Gamma}_\textrm{min}(z)$, the filter gains can be adjusted based on the ratio $\alpha = m_i / m_\textrm{min}$~\cite{valimaki2024two}. 

\subsection{Frequency Sampling for Differentiable Filters}
Several techniques are available for implementing differentiable linear and time-invariant (LTI) filters. Finite impulse response (FIR) filters lend themselves naturally to convolutional-layer implementations~\cite{hayes2024review, mezza2024modeling}, whereas infinite impulse response (IIR) filters are more challenging because of their recursive structure and inherent stability requirements. A common approach is to implement IIR filters in the time domain using recurrent neural networks and backpropagation through time, although this strategy often results in high memory consumption. To alleviate the limitations of time-domain optimization~\cite{hayes2024review, kuznetsov2020differentiable, steinmetz2022style}, the frequency sampling method provides an effective alternative. In this approach, an FIR approximation of the desired filter is obtained by sampling its frequency response at selected complex frequencies~\cite{nercessian2020neural, nercessian2021lightweight}. The sampling resolution is chosen to minimize time-aliasing distortion~\cite{lee2022differentiable, nercessian2021lightweight}.
 
The response of a real-valued LTI filter is sampled on a vector of linearly-spaced frequencies 
from 0 to $\pi$ rad/sample:
\begin{align}\label{eq:freq_vector}
    \boldsymbol{z}_{M} = [e^{\imath \pi \frac{0}{M-1}}, e^{\imath \pi \frac{1}{M-1}}, \dots, e^{\imath \pi \frac{M-2}{M-1}},  e^{\imath \pi }],
\end{align}
where $M$ is the number of frequency bins and $\imath = \sqrt{-1}$. 
The frequency response of a  SISO IIR filter $H(z)$ of order $\mathcal{{M}}$ can be evaluated over $\boldsymbol{z}_M$ as follows 
\begin{align}\label{eq:freqz}
H(\boldsymbol{z}_M) & = \frac{\text{DFT}(\mathbf{b})}{\text{DFT}(\mathbf{a})} \nonumber \\ &= \frac{{b}_{H, 0}+\textup{b}_{H,1}\boldsymbol{z}_M^{-1}+\dots+\textup{b}_{H,\mathcal{{M}}}\boldsymbol{z}_M^{-\mathcal{{M}}+1}}{\textup{a}_{H,0}+\textup{a}_{H,1}\boldsymbol{z}_M^{-1}+\dots+\textup{a}_{H,\mathcal{{M}}}\boldsymbol{z}_M^{-\mathcal{{M}}+1}}, \\
\text{where}\\
\mathbf{b} &= [b_{H, 0}, b_{H, 1}, \dots, b_{H, \mathcal{{M}}}]\\
\mathbf{a} &= [a_{H, 0}, a_{H, 1}, \dots, a_{H, \mathcal{{M}}}]
\end{align}

A drawback of the frequency sampling methods is time-aliasing, which becomes relevant when the effective duration $L$ of the filter impulse response (IR) exceeds the Fourier transform length $2(M-1)$~\cite{smith2007math, lee2022differentiable, hayes2024review}. This issue is especially significant for lossless systems, where certain discrete frequency points cause the transfer function to diverge to infinity.
For lossy IIR filters, the time-aliasing error asymptotically decreases as $M$ increases~\cite{lee2022differentiable}, at the cost of higher computational cost. In~\cite{santo2024flamo}, the authors presented a method to mitigate distortions by sampling the transfer function outside the unit circle.

\subsection{Loss Profile}\label{subsec:loss_land}
Whether the filter parameters are learned via simple gradient-based optimization or via a neural network, the underlying problem typically involves minimizing a high-dimensional, non-convex loss function by adjusting a set of parameters $\theta$, which may represent either the neural network weights or the DDSP system parameters. The convergence of such methods depends on the quality of the loss profile, which in turn is influenced by various design choices and modeling considerations~\cite{li2018visualizing}.
A straightforward approach to visualize a loss function is to select two parameter sets, $\theta'$ and $\theta''$, such that the target parameters $\theta^*$ lie on the line interpolating between them, and then plot the loss values along this line~\cite{li2018visualizing}. This technique provides insight into the sharpness of different minima and, at higher resolutions, the smoothness of the overall loss profile.
However, visualization is feasible only when $\theta$ is at most two-dimensional, which limits its applicability to higher-dimensional loss functions and prevents the direct observation of non-convexities. Despite this constraint, in DDSP applications, this method has proven valuable for evaluating the quality of loss functions, particularly for sinusoidal frequency estimation tasks~\cite{schwar2023multi, hayes2023sinusoidal}. For more advanced visualization techniques, the reader is referred to the work of Li et al.~\cite{li2018visualizing}. 

\subsection{Loss Functions}\label{subsec:loss_fn}
The loss functions used in this work are based on dissimilarity measures between representations of the system's IR, $\hat{h}(t)$, and the target RIR, $h(t)$, 
\begin{align}\label{eq:general_loss}
\mathcal{L} = \mathcal{J}\Bigl(q\bigl({h}(t)\bigr), q\bigl(\hat{h}(t)\bigr)\Bigr)~,
\end{align}
where $\mathcal{J}(\cdot)$ denotes a chosen dissimilarity measure, and $q(\cdot)$ represents a transformation applied to the IRs, such as the short-time Fourier transform (STFT) or the EDC. 
This approach corresponds to a learning paradigm that does not rely on knowledge of either the ground-truth $\Tsixty$ or the attenuation filter parameters. 

The MSS loss is a widely used loss function in audio machine learning, with applications across numerous tasks, including those related to reverberation~\cite{steinmetz2021filtered, lee2022differentiable, bona2022automatic, su2020acoustic, gotz2025matching}. MSS addresses the inherent trade-off between time-frequency resolution present in magnitude spectrograms by incorporating multiple STFTs with varying time-frequency resolutions into a unified loss function~\cite{yamamoto2020parallel}.
The MSS is composed of a spectral convergence term $\mathcal{L}_{\text{SC}}$ and a spectral log-magnitude term $\mathcal{L}_{\text{SM}}$, respectively:
\begin{align}\label{eq:mss_sc}  
\mathcal{L}_{\text{SC}}(h, \hat{h}) &= \frac{\lVert|{H}(t,f)|-|{\hat{H}(t,f)}|\rVert_F}{\lVert|{H}(t,f)|\rVert_F}
\end{align}
and
\begin{align}\label{eq:mss_sm}
\mathcal{L}_{\text{SM}}(h, \hat{h}) = \frac{1}{S} \lVert&\log(|{H}(t,f)|)-\log(|{\hat{H}(t,f)}|)\rVert_1\,,
\end{align}
where $\lVert\cdot\rVert_\textrm{1}$ is the $\ell_1$ norm, $\lVert\cdot\rVert_F$ is the Frobenius norm, $|\cdot|$ is the absolute value, and $S$ is the number of STFT frames.
The MSS loss is defined as the average error across each of the $R$ resolutions, i.e.,
\begin{align}\label{eq:yamamoto_mss}
    \mathcal{L}_{\text{MSS}}(h, \hat{h})  = \frac{1}{R}\sum_{r=1}^{R} (\mathcal{L}_{\text{SC}}^{(r)}(h, \hat{h}) + \mathcal{L}_{\text{SM}}^{(r)}(h, \hat{h}))\,.
\end{align}

To achieve optimal performance, it is necessary to select appropriate values for the frame size, window type, and hop size, as systematic analyses have shown that different hyperparameter configurations can significantly influence the loss. This is particularly evident in tasks such as sinusoid frequency estimation, where spectral leakage can lead to periodic fluctuations in the loss profile~\cite{schwar2023multi}. For this work, however, we assume that the effect of spectral leakage on RIRs is negligible, and validated our choice of hyperparameters by visual inspection of the representations in Sec.~\ref{subsec:representations}.

A more common representation in artificial reverberation is the EDC, which is used to calculate $\Tsixty$~\cite{schroeder1965new}. In its original definition, the EDC is computed over the full spectrum of the RIR. In this paper, we generalize the EDC to multiple frequency bands in order to capture frequency-dependent $\Tsixty$. 
Given $h(t)$ of length $L$ samples, the EDC for a specific frequency band can be computed using Schroeder backward integration: 
\begin{align}\label{eq:edc_schroeder}
    \varepsilon(t; f_\textrm{bn}) = \sum_{\tau=t}^Lh_{f_\textrm{bn}}^2(\tau)\,,
\end{align}
where $h_{f_\textrm{bn}}$ is the input RIR at a frequency band with center frequency $f_\textrm{bn}$. It is usually reported on a dB scale, here denoted by $\varepsilon_{\textrm{dB}}$. The EDC, with its frequency-dependent variations, has been previously presented as loss function in~\cite{mezza2024data, ratnarajah2022mesh2ir, dal2024similarity}, with general for
\begin{align}\label{eq:edc_loss}
    \mathcal{L}_{\text{EDC, dB}} = \frac{\sum_{f_\textrm{bn} \in \mathcal{B}}\sum_{t=0}^L ( \varepsilon_{\textrm{dB}}(t; f_\textrm{bn}) - \hat{\varepsilon}_{\textrm{dB}}(t; f_\textrm{bn}))^2}{\sum_{f_\textrm{bn} \in \mathcal{B}}\sum_{t=0}^L \varepsilon_{\textrm{dB}}(t; f_\textrm{bn})^2},
\end{align}
where $\mathcal{B}$ indicates the set of frequency bands. The term in the denominator normalizes the loss with respect to the target EDC energy, stabilizing its dynamic range and making it more suitable for use in composite loss functions. This normalization rescales the gradient magnitude without altering its direction.

Coloration and sparsity artifacts in delay networks can degrade their perceptual quality even when attenuation filters are perfectly matched to the target RIR. To address these artifacts, the authors in \cite{dal2023differentiable, santo2024optimizing} proposed two loss functions that iteratively minimize coloration while increasing temporal density for a given set of delay lengths. The latter, referred to as sparsity loss $\mathcal{L}_U$, encourages a fast buildup of temporal reflection, resulting in a perceptually smooth reverberation tail~\cite{dal2024rir2fdn}. In Sec.~\ref{subsec:grad_optim} $\mathcal{L}_U$ is included in combination with the aforementioned loss functions, using the following definition:
\begin{equation} \label{eq:sparsity_loss}
    \mathcal{L}_U=\frac{N\sqrt{N}-\sum_{i,j}\lvert U_{i,j}\rvert}{N\left(\sqrt{N}-1\right)},
\end{equation}
where $U_{i,j}$ is the entry of matrix $\boldsymbol{U}$ in the $i$-th row and $j$-th column.


\section{METHOD}\label{sec:method}
For the study presented in this paper, it is important to note that the system described in Eq.~\eqref{eq:tr_fdn} does not model any sources of noise, including background or measurement noise. As a result, when FDNs are optimized on measured RIRs, a model mismatch occurs. Analyzing the effect of this mismatch on the loss profile is therefore crucial to assess the errors it introduces.

In this section, we introduce a method for visualizing loss profiles and assessing their variability under noisy targets and additional optimization objectives, such as minimizing coloration and controlling temporal density. The loss functions used in the rest of the paper are introduced in Sec.~\ref{subsec:loss_fn}. 

\subsection{First Order Shelving Filter}
\begin{figure}[!t]
    \centering
    \includegraphics[trim={0.25cm 0.25cm 0.8cm 0cm}, clip, width=0.5\linewidth]{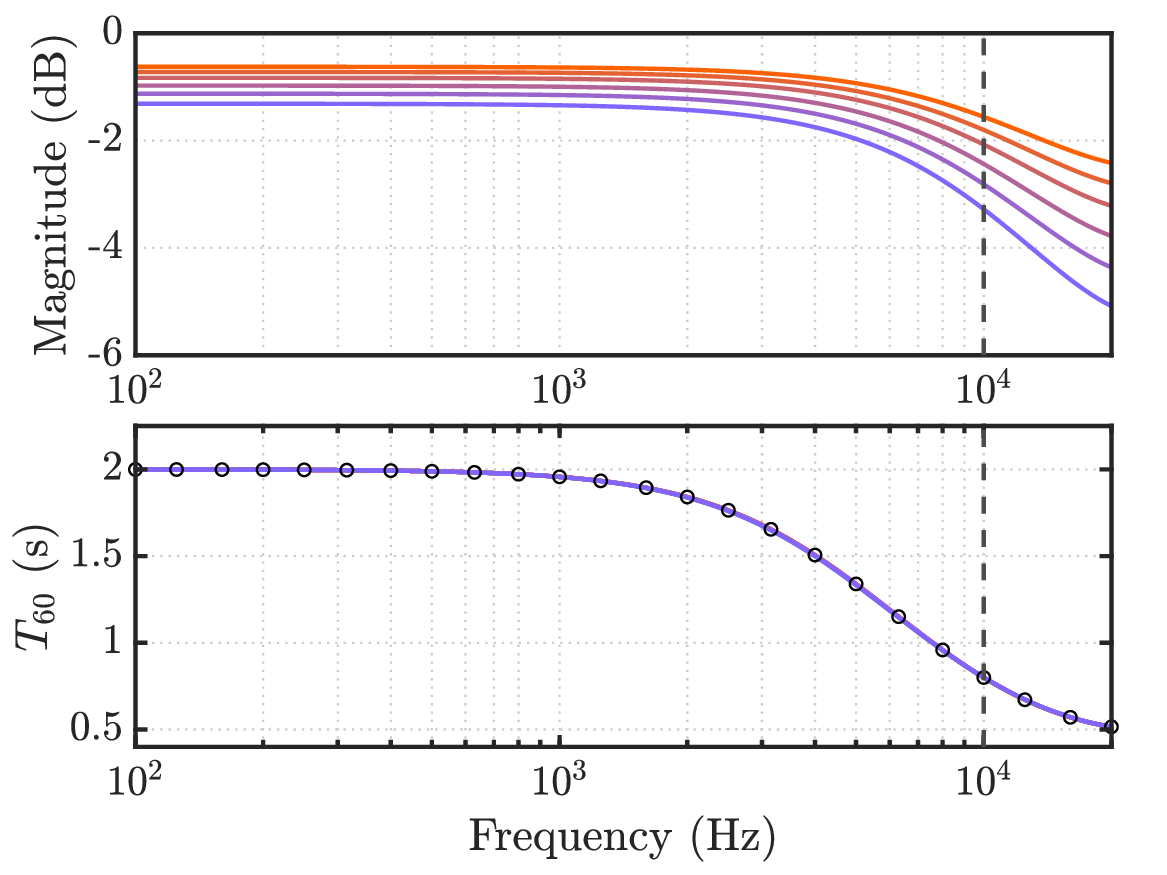}
    \caption{Magnitude response (top) of the first-order shelving filter for each delay line with lengths [997, 1153, 1327, 1559, 1801, 2099] samples, and the corresponding $T_{60}$ curve (bottom) obtained by solving Eq.~\eqref{eq:scaled_filters}. The shortest and longest delay lines are indicated in orange and purple colors, respectively. The vertical dashed lines indicate the crossover frequency $f_\textrm{c} = 10$~kHz.}
    \label{fig:prop_shelving}
\end{figure}
In this study, we use a first-order shelving filter to design $\mat{\Gamma}(z)$. This simple filter structure allows us to visually inspect the loss profile across two parameters: the reverberation time at dc, $T_{60}^{\textrm{dc}}$, and the crossover frequency $f_\textrm{c}$. In~\cite{jot2015proportional}, Jot presented a proportional shelving filter with a mutually homothetic magnitude response for varying dc gains. We adapt the original design to achieve a nonzero dB gain at the Nyquist limit while maintaining an approximately proportional response.

The following parameters govern the modified shelving filter's response:
\begin{equation}
\begin{aligned}
g &= \frac{\gamma_{\text{dc}}}{\gamma_{\text{Ny}}}~, \quad
\omega_\textrm{c} = \frac{2\pi f_\textrm{c}}{f_\textrm{s}}~
\end{aligned}
\end{equation}
where $\gamma_{\text{dc}}$ and $\gamma_{\text{Ny}}$ represent the target filter linear gains at dc and Nyquist limit, respectively. Using these parameters, the filter's transfer function can be written as
\begin{equation}
{\Gamma}(z)=\frac{{b}_{\Gamma,0}+{b}_{\Gamma,1}z^{-1}}{{a}_{\Gamma,0}+{a}_{\Gamma,1}z^{-1}},
\label{eq:shelf}
\end{equation}
\noindent where the coefficients are
\begin{equation}
\begin{aligned}\label{eq:low-shelving}
b_{\Gamma,0} &= \gamma_{\text{Ny}} (\sqrt{g} \tan(\omega_\textrm{c}) + 1)~, \\
b_{\Gamma,1} &= \gamma_{\text{Ny}} (\sqrt{g} \tan(\omega_\textrm{c}) - 1)~, \\
a_{\Gamma,0}&= \frac{\tan(\omega_\textrm{c}) }{\sqrt{g}} + 1~, \quad
a_{\Gamma,1} = \frac{\tan(\omega_\textrm{c}) }{\sqrt{g}} - 1~.
\end{aligned}
\end{equation}

The top pane of Fig.~\ref{fig:prop_shelving} shows the magnitude responses of the resulting filters $\Gamma(z)$, each scaled by a different delay-line length, i.e., $|\Gamma_i| = |\Gamma|^{m_i}$ for $i = 1, \dots, 6$. The reverberation times at dc and Nyquist limit are $T_{60}^{\textrm{dc}} = 2$\,s and  $T_{60}^{\textrm{Ny}} = 0.5$\,s, respectively, while the sample rate is $\fs = 48$\,kHz. The bottom panel in Fig.~\ref{fig:prop_shelving} shows the corresponding $T_{60}$ curve to which each scaled filter maps. The curves overlap almost perfectly, indicating that the filters are approximately proportional and yield the same decay rate.

\subsection{Simulating Parameter Perturbation}\label{subsec:parameters_perturbation}

When minimizing loss functions of the form in Eq.~\eqref{eq:general_loss}, the effect of the attenuation filters on the output cannot be fully isolated from the contribution of the frequency-independent parameters $\theta_{\text{FI}} = \{\boldsymbol{U}$, $\boldsymbol{b}$, $\boldsymbol{c}$\}. Although the modal decay rates are determined uniquely by $\boldsymbol{\Gamma}(z)$, these parameters still influence modal characteristics~\cite{schlecht2015time, schlecht2019modal}, as discussed in Sec.~\ref{subsec:param_role}. One way to stabilize the system's response during optimization would be to fix the frequency-independent parameters. However, this approach precludes joint minimization with losses such as Eq.~\eqref{eq:sparsity_loss} and may, intuitively, lead to convergence to undesirable local minima.

\begin{figure}[!t]
    \centering
    \includegraphics[trim={1.15cm 0.5cm 1.25cm 0.3cm}, clip, width=0.75\linewidth]{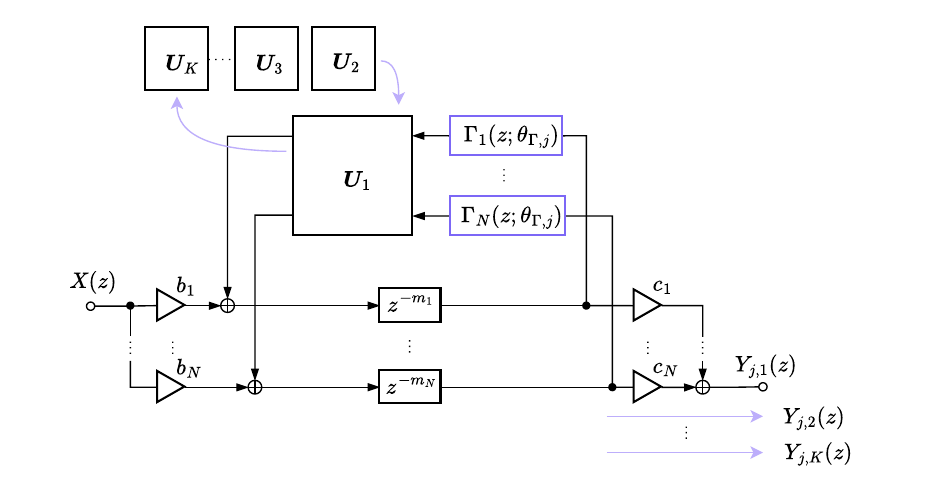}
    \caption{Schematic example of parameter perturbations applied to the FDN. The output of the system for each attenuation parameter state $\theta_{\Gamma,j}$ is computed for different instances of the mixing matrix $\mat{U}_k$, where $k$ is the perturbation index, producing output $Y_k(z)$}.
    \label{fig:fdn_pert}
\end{figure}

To verify the intuition described above and to examine how auxiliary losses affect Eq.~\eqref{eq:general_loss} through perturbations of the parameters in $\theta_{\text{FI}}$, we compute the loss profile at different instances of $\theta_{\text{FI}}$. Fig.~\ref{fig:fdn_pert} illustrates the proposed approach, using $\boldsymbol{U}$ as an example. We use the method introduced in Sec.~\ref{subsec:loss_land} to compute the loss profile between two states, $\theta_{\mat{\Gamma}}'$ and $\theta_{\mat{\Gamma}}''$, such that the target values $\theta^*_{\mat{\Gamma}}$ fall within the interval $[\theta_{\mat{\Gamma}}', \theta_{\mat{\Gamma}}'']$. This interval is divided into linearly or logarithmically spaced points, denoted $\theta_{\mat{\Gamma}, j}$, where $j$ indicates the step index. At each step $j$, the FDN's IR is computed for $K$ different instances of a chosen parameter, such as $\boldsymbol{U}$, whose value is drawn from a normal distribution with a constant mean and standard deviation across instances. For $\boldsymbol{U}$, these random values are mapped to the orthogonal space before the forward path. Each instance produces an IR given by
\begin{align}
h_{j,k} = \text{IDFT}\left({H_{j,k}}\right) = \text{IDFT}\left(\frac{Y_{j,k}(z)}{X(z)}\right)\,,
\end{align}
where IDFT indicates the inverse discrete Fourier transform, and
\begin{align}
H_{j,k}(z) = \mat{c}^\top\big[\mat{D_m}(z)^{-1} -\mat{U}_k\mat{\Gamma}(z; \theta_{\Gamma,j})\big]^{-1}\mat{b}\,,
\end{align}
which is then used to compute the loss values against a target IR generated from a random instance of a different FDN with target attenuation parameters $\theta^*_{\mat{\Gamma}}$. This experiment's configuration details and results are presented in Sec.~\ref{subsec:param_perturb}.

\subsection{Explorative Analysis}\label{subsec:representations}
Before running the gradient-descent optimization, we evaluate the FDN constructed with the target filters $\mat{\Gamma}(z)$ to ensure that the chosen loss function is likely to exhibit a meaningful minimum. To assess this before computing the loss profile, we visualize the pairwise distances in both the STFT and EDC domains, computed over time-frequency bins and time bins, respectively, between the target $h(t)$ and the IR of the learnable FDN, $\hat{h}(t; \theta^*_{\Gamma})$. We generate the target late reverberation $h(t)$ using an FDN of size $N=32$, while $\hat{h}$ is generated using an FDN of size $N=6$. The attenuation filter follows the design in Fig.~\ref{fig:prop_shelving}. Both FDNs are initialized with random orthogonal $\mat{U}$, $\mat{c}=\mat{1}$, and $|\mat{b}|=\mat{1}$ such that entries of $\mat{b}$ have alternate signs. To highlight the effect of background noise, we add white Gaussian noise $w(t)$ to $h(t)$ at an SNR of 70\,dB. 

\begin{figure}[!t]
    \centering
    \includegraphics[trim={0cm 0cm 0cm 0cm}, clip, width=0.75\linewidth]{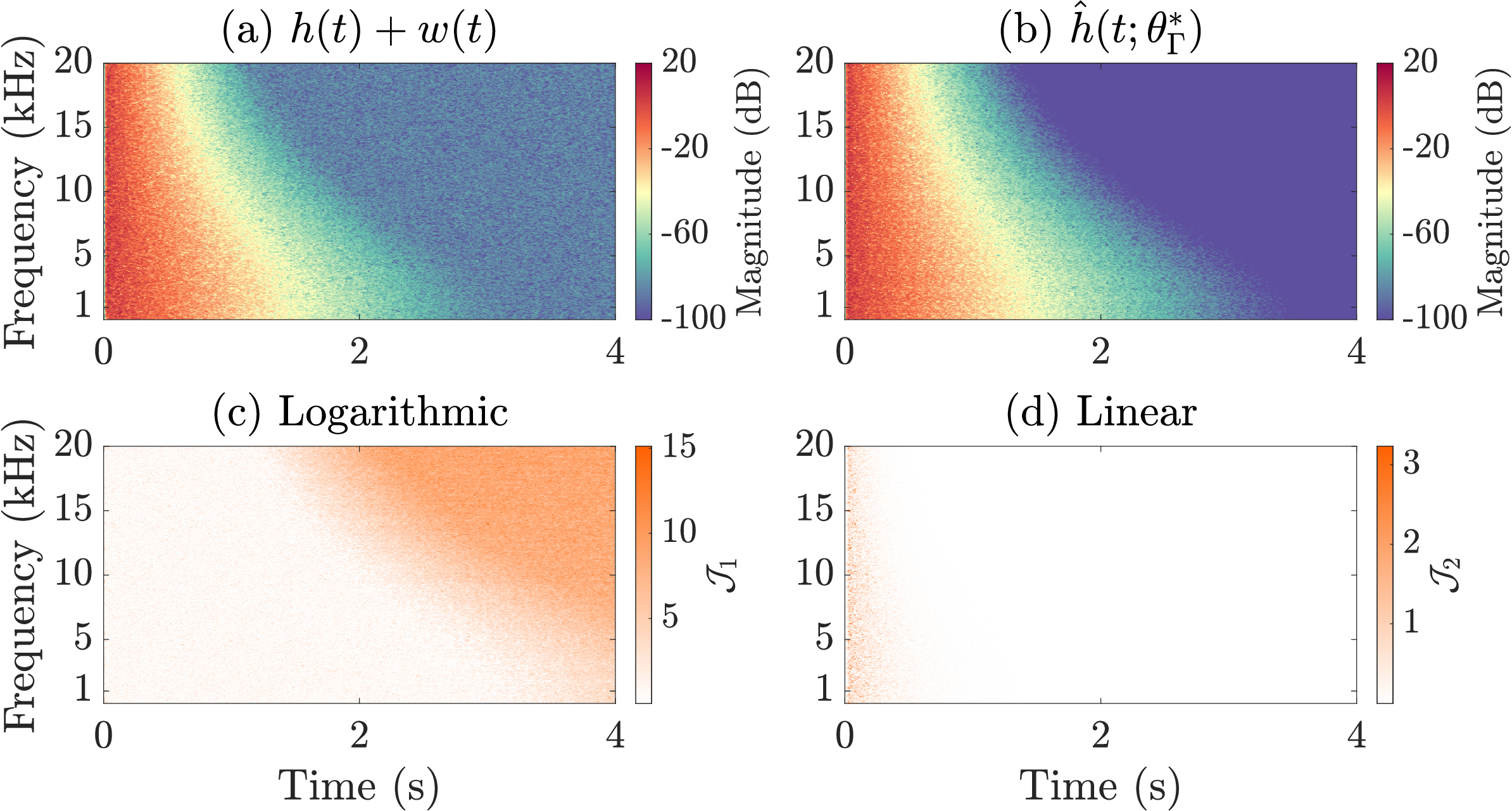}
    \caption{STFT of (a) the noisy synthetic target and (b) the modeled IR $\hat{h}(t; \theta^*_{\Gamma})$, and their difference in (c) logarithmic and (d) linear scale according to Eq.~\eqref{eq:log_diff} and Eq.~\eqref{eq:lin_diff}, respectively.}
    \label{fig:stft_plots}
\end{figure}

Fig.~\ref{fig:stft_plots} illustrates the STFT, computed with a window size of 2048 samples and hop size of 512 samples,  of (a) $h(t)$ and (b) $\hat{h}(t; \theta^*_{\Gamma})$, along with the distance measures used in Eq.~\eqref{eq:yamamoto_mss}: 
\begin{align}    
\mathcal{J}_{1}(h, \hat{h}) &= |\log(|H(t,f)|)-\log(|\hat{H}(t,f)|)| \label{eq:log_diff} \\
\mathcal{J}_{2}(h, \hat{h}) &= ||H(t,f)|-|\hat{H}(t,f)|| \label{eq:lin_diff}
\end{align}

The background noise is clearly captured in Fig.~\ref{fig:stft_plots}~(a), where the decay is interrupted by a homogenous noise floor at approximately -100~dB. The response in Fig.~\ref{fig:stft_plots}~(b) shows light resonances due to a lower number of modes and thus a sparser spectrum compared to~(a). 
The $\mathcal{J}_{\text{1}}$ metric significantly penalizes differences in the background noise, as seen in Fig.~\ref{fig:stft_plots}~(c). This feature can introduce instabilities during training, particularly when the noise is not incorporated into the reverb model. In contrast, the error in $\mathcal{J}_{\text{2}}$, shown in Fig.~\ref{fig:stft_plots}~(d), is not affected by the background noise, and shows errors only until the echo density saturates. 

\begin{figure}[!t]
    \centering
    \includegraphics[trim={0cm 0cm 0cm 0cm}, clip, width=0.75\linewidth]{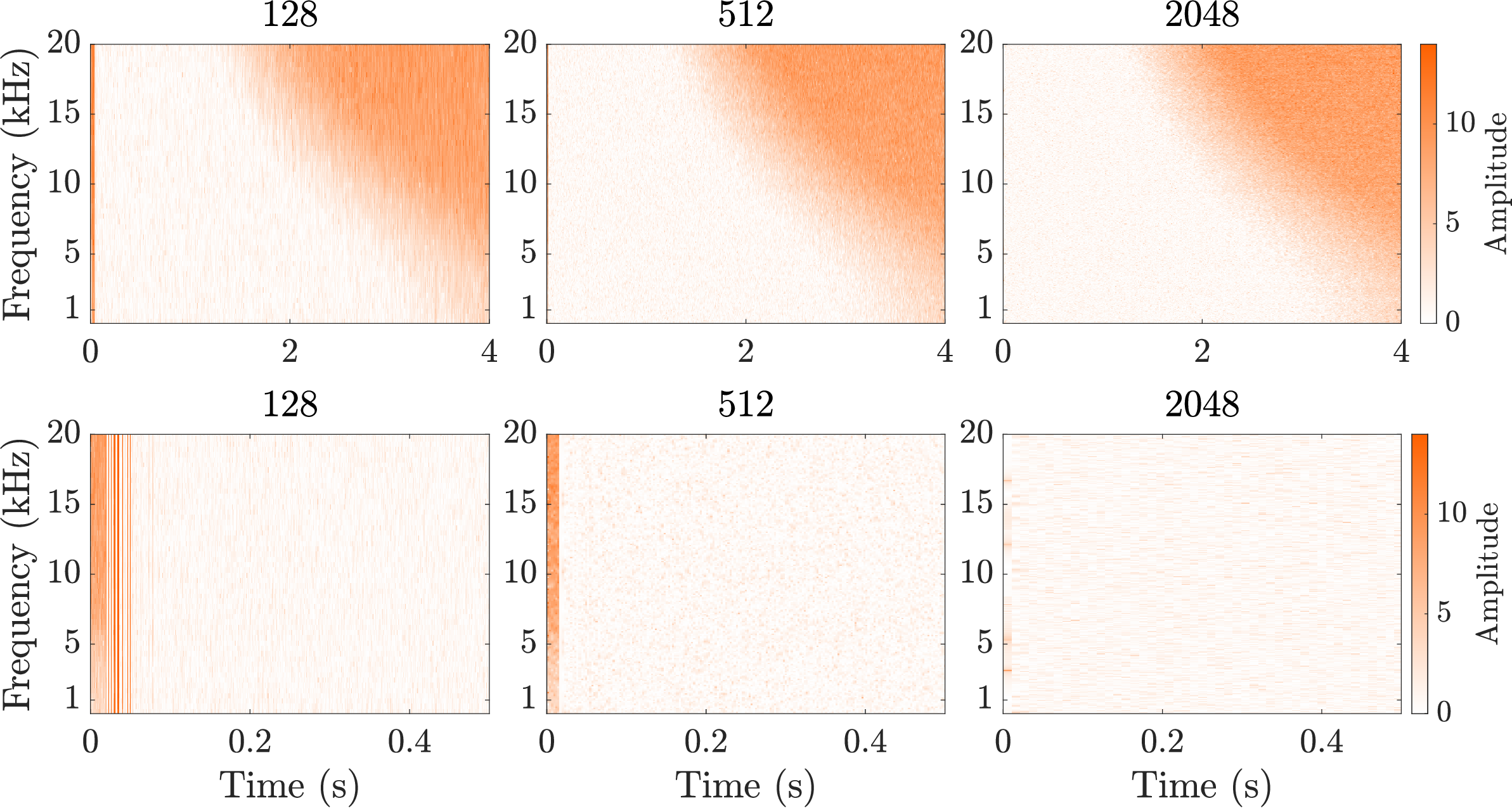}
    \caption{(top) Distance $\mathcal{J}_{1}$ between the STFT target $h(t)$ and modeled IR $\hat{h}(t;\theta^*_{\Gamma})$ for different window lengths.
(bottom) Zoomed view of the distance over the first 500 ms}
    \label{fig:mss_log}
\end{figure}

\begin{figure}[!t]
    \centering
    \includegraphics[trim={0cm 0cm 0cm 0cm}, clip, width=0.75\linewidth]{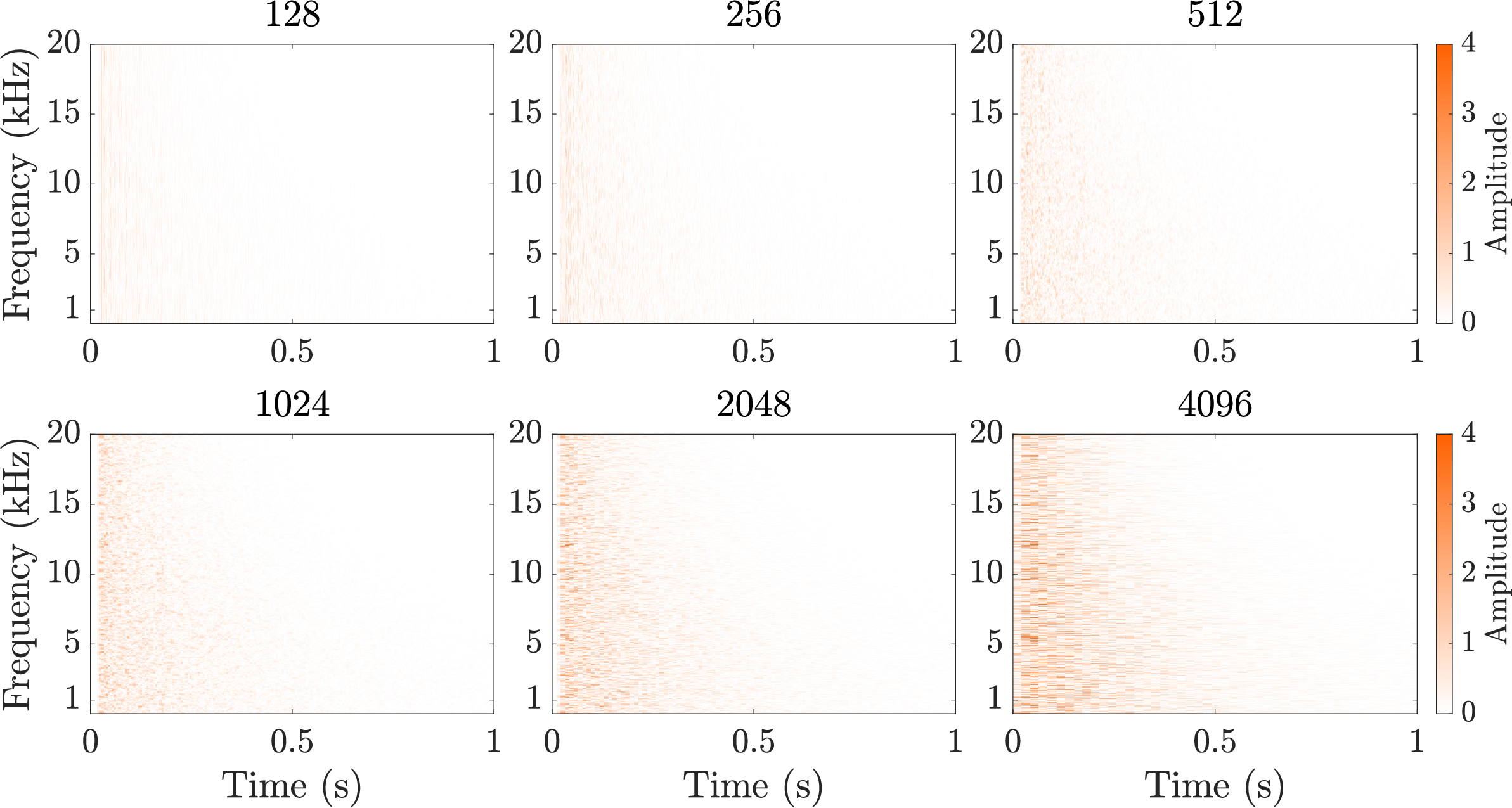}
    \caption{Linear distance $\mathcal{J}_{\text{2}}$ between the STFT target $h(t)$ and modeled IR $\hat{h}(t; \theta^*_{\Gamma})$ at different window lengths between 128 (top left) and 4096 (bottom right) samples.}
    \label{fig:mss_lin}
\end{figure}

Figure~\ref{fig:mss_log} shows $\mathcal{J}_1$ for three different window sizes: 128, 512, and 2048 samples. The top panels show that reaching the noise floor in $h(t)$ leads to high error values. The bottom panels highlight the large errors in the early frames, caused by differences in the buildup of echo density in the onset region. These differences are particularly pronounced for shorter window lengths. Fig.~\ref{fig:mss_lin} shows a zoomed view of $\mathcal{J}_2$ for different window sizes ranging from 128 to 4096 samples. For larger window sizes, which provide higher frequency resolution, the error is larger. In contrast to $\mathcal{J}_1$, the linear distance $\mathcal{J}_2$ does not capture the difference in noise floor level.  

The EDC, computed according to Eq.~\eqref{eq:edc_schroeder}, describes the remaining signal energy from each time index onward. To split the signal into subbands, we use an energy-preserving fractional-octave filter bank\footnote{\url{https://pyfar.readthedocs.io/en/stable/modules/pyfar.dsp.filter.html}}, which does not introduce significant delay. When comparing EDCs at each time index, several factors must be considered. The first value, $\varepsilon(0; f_\textrm{bn})$, represents the total energy of the signal within the analyzed frequency band. Jointly estimating both $\varepsilon(0; f_\textrm{bn})$ and $T_{60}(f_\textrm{bn})$ can lead to a more irregular loss profile due to more degrees of freedom in the optimization. In particular, for the case of the MSS loss, mismatches in overall energy can be partially compensated for by adjusting $f_\textrm{c}$ and $T_{60}^{\textrm{dc}}$ towards lower values. While in the case of EDC distances, energy mismatches can produce intersecting curves and consequently introduce local minima. To mitigate this, we match the energy of $\hat{h}(t; \theta^*_{\Gamma})$ to that of $h(t)$ by appropriately scaling the input and output gains.

\begin{figure}[!t]
    \centering
    \includegraphics[trim={0cm 0.25cm 0cm 0cm}, clip, width=0.75\linewidth]{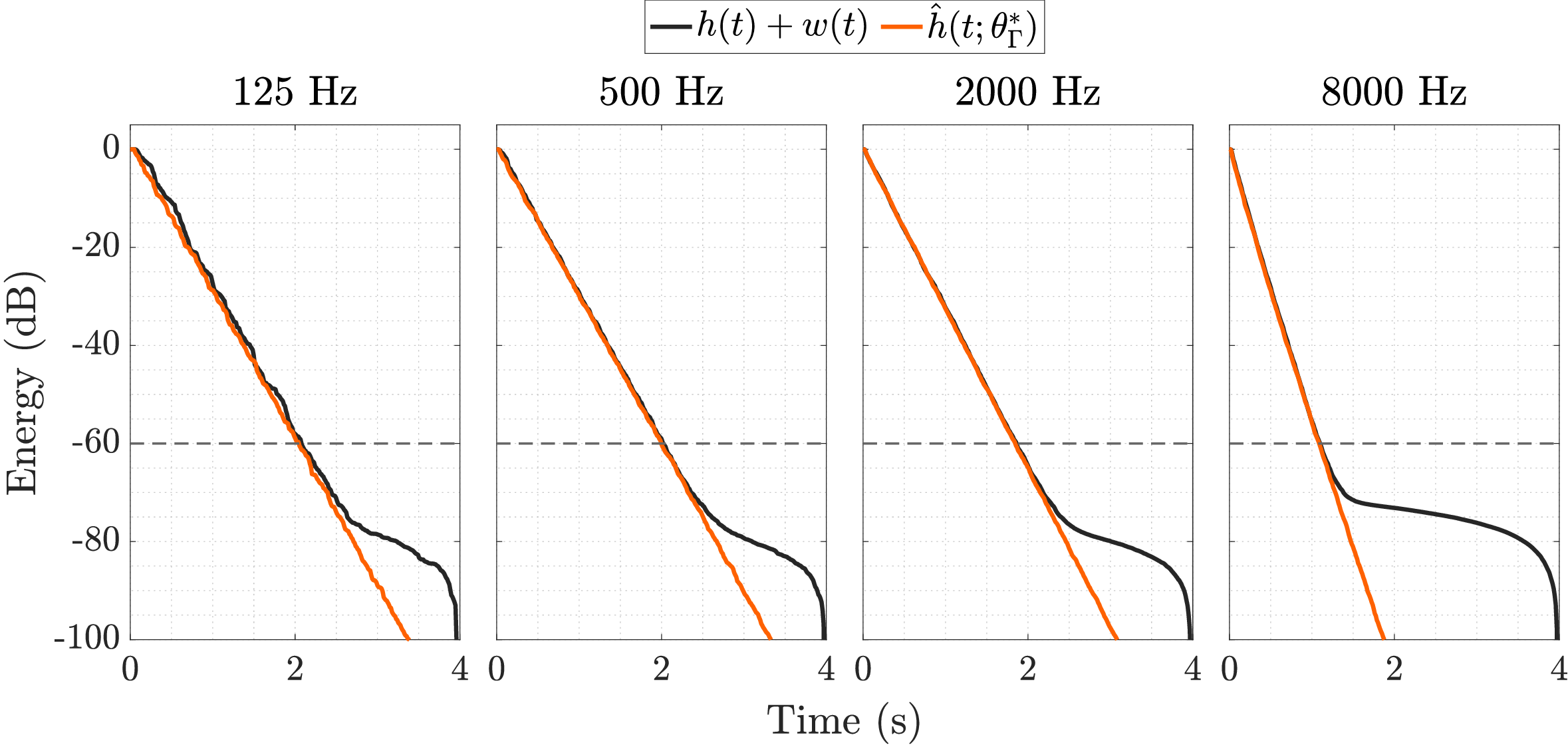}
    \caption{EDC of the of target $h(t)$and modeled IR $\hat{h}(t; \theta^*_{\Gamma})$ at four one-octave bands. Dashed lines indicate the -60~dB level.}
    \label{fig:edc_curves}
\end{figure}

Figure~\ref{fig:edc_curves} shows the EDCs of $h(t)$ and $\hat{h}(t; \theta^*_{\Gamma})$ at four one-octave bands. The values are correctly aligned at $t = 0$, and the decay behavior is also well matched. Before reaching the noise floor, the curves differ slightly, mainly due to differences in echo and modal density, which are more noticeable in the lower frequency bands. The target signal’s background noise appears as a clear plateau, whereas $\hat{h}(t; \theta^*_{\Gamma})$ exhibits a clean decay.

\subsection{Effect of Noise on the Loss Profile}\label{subsec:noise_loss_profile}
\begin{figure*}[!h]
    \centering
    \includegraphics[width=\linewidth]{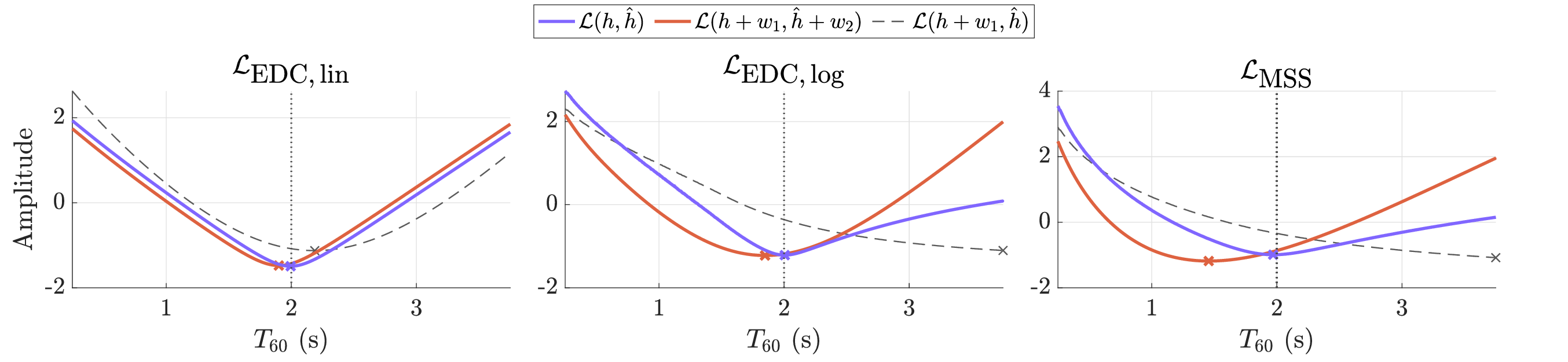}\vspace{0.2cm}
    \includegraphics[trim={0 0 0 1.25cm}, clip, width=\linewidth]{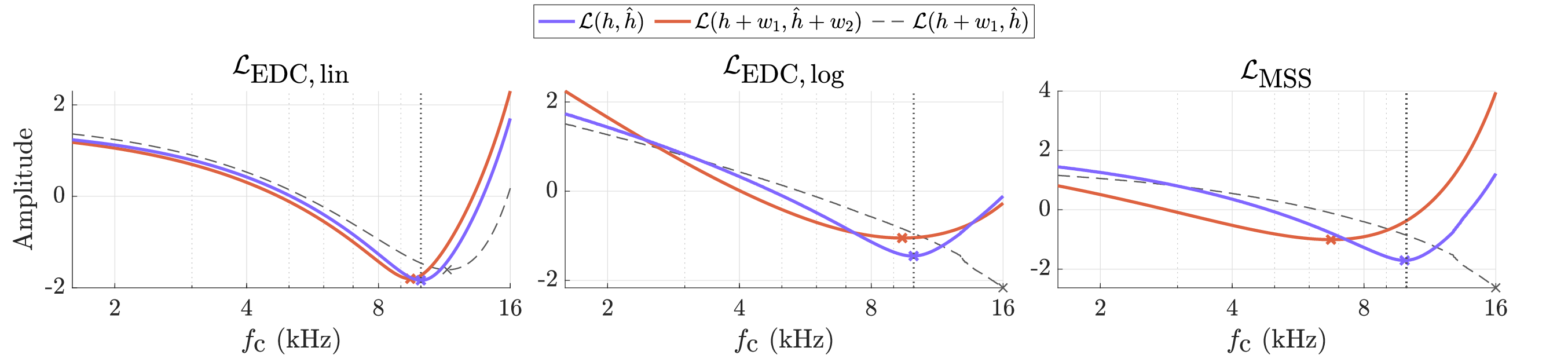}
    \caption{Loss profile for 1000 steps between two states of $\theta_{\mat{\Gamma}} = (\Tsixty^{\textrm{dc}}, f_\textrm{c})$. (top) The loss profile is computed in the range $[\theta_{\mat{\Gamma}}' = (0.25~\text{s}, f_\textrm{c}^*), \theta_{\mat{\Gamma}}'' = (3.75~\text{s}, f_\textrm{c}^*)]$, and with target $\Tsixty^{\textrm{dc}~*} = 2$~s. The curve minimum is indicated with $\times$. (bottom) The loss profile is computed in the range $[\theta_{\mat{\Gamma}}' = (\Tsixty^{\textrm{dc}~*}, 1.6~\text{kHz}), \theta_{\mat{\Gamma}}'' = (\Tsixty^{\textrm{dc}~*}, 16~\text{kHz})]$, and with target $f_\textrm{c}^* = 10$~kHz. Targets are indicated with a vertical dotted line. The curve minimum is indicated with $\times$. For ease of comparison, all loss curves were standardized to zero mean and unit standard deviation.}
    \label{fig:loss_profile_I}
\end{figure*}

The final step in the analysis evaluates the accuracy and smoothness of the selected representations by computing their loss profiles. Fig.~\ref{fig:loss_profile_I} presents the loss profiles as functions of $\Tsixty^{\textrm{dc}}$ (top) and $f_\textrm{c}$ (bottom). We consider three different losses: the $\mathcal{L}_{\textrm{EDC}}$ loss defined in Eq.~\eqref{eq:edc_loss}, computed using either the linear-scale EDC ($\mathcal{L}_{\textrm{EDC,\,lin}}$) or the logarithmic-scale EDC ($\mathcal{L}_{\textrm{EDC,\,log}}$), and the $\mathcal{L}{_\textrm{MSS}}$ loss defined in Eq.~\eqref{eq:yamamoto_mss}. All losses exhibit a minimum at the target parameter values, indicated by the $\times$ markers. For the EDC-based losses, we used EDCs in one-octave bands from 31.5 Hz to 16 kHz, while the STFT used in $\mathcal{L}_{\textrm{MSS}}$ is computed at six resolutions, $R = {128, 256, 512, 1024, 2048, 4096}$ with 75\% overlap and Hann window.

To analyze the effect of background noise, Fig.~\ref{fig:loss_profile_I} shows loss profiles under two additional conditions: when noise is present in the target only, $\mathcal{L}(h + w_1, \hat{h})$, and when noise is added to both signals, $\mathcal{L}(h + w_1, \hat{h} + w_2)$, at SNR=10 dB. We refer to the latter case as the \textit{noise-aware} condition. In the first case, the minimum shifts toward higher parameter values, with $\mathcal{L}_{\textrm{MSS}}$ and $\mathcal{L}_{\textrm{EDC,\,log}}$ showing the largest deviations from the noiseless condition. When noise is also added to $\hat{h}(t)$, the location of the minimum and the overall trend of $\mathcal{L}_{\textrm{EDC}}$ become more closely aligned with the noiseless case. Although it improves in the noise-aware case, $\mathcal{L}_{\textrm{MSS}}$ still exhibits a substantial error. The mean absolute error (MAE) between the attenuation parameters at the loss minima and the target parameters, under the noise-aware condition, is reported in the first rows of Table~\ref{tb:mae_perturbation}. Two-dimensional plots showing local convexity of the loss surface over $T_{60}$ and $f_\textrm{c}$ are provided in the \hyperref[appendix_surface]{Appendix}.

\begin{figure*}[!h]
    \centering
    \includegraphics[width=\linewidth]{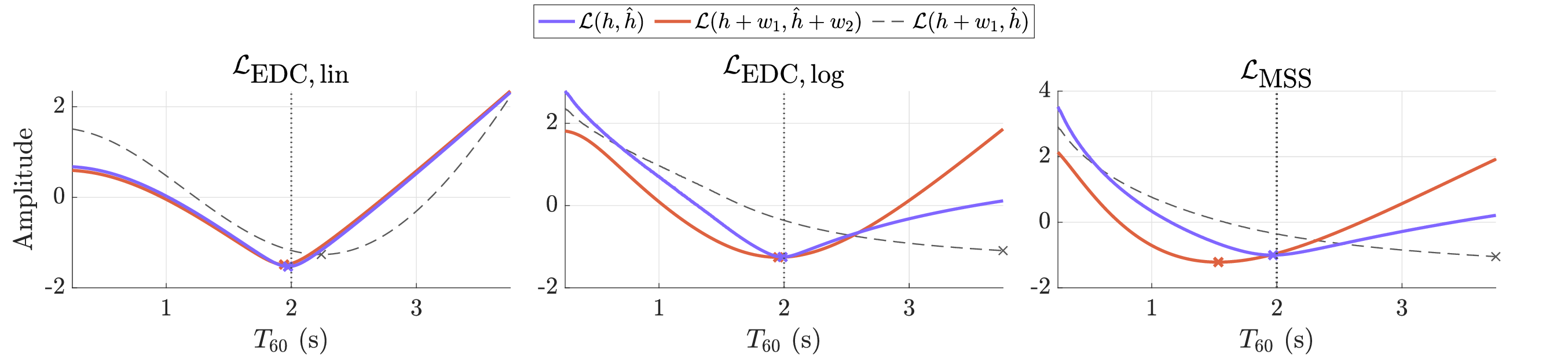}\vspace{0.2cm}
    \includegraphics[trim={0 0 0 1.25cm}, clip, width=\linewidth]{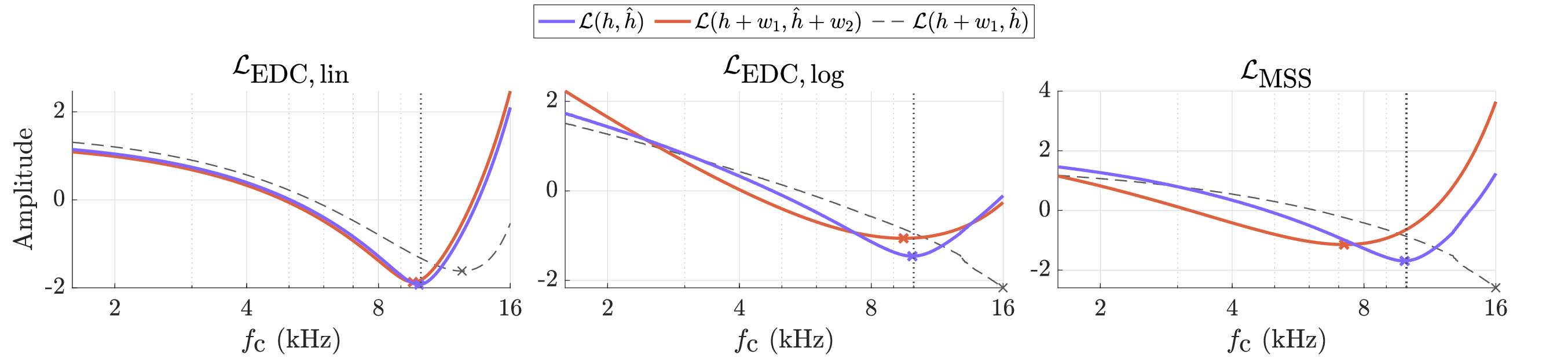}
    \caption{Loss profile for 1000 steps between two states of $\theta_{\mat{\Gamma}} = (\Tsixty^{\textrm{dc}}, f_\textrm{c})$, analogous to Fig.~\ref{fig:loss_profile_I}. Before computing the loss, the early reflections in both FDN IRs were discarded up to $t_\textrm{mix} = 87.5$ ms. }
    \label{fig:loss_profile_II_edc}
\end{figure*}

We perform a similar test after removing the first $t_{\textrm{mix}}$\,s from both the analyzed and target responses, where $t_{\textrm{mix}}$ denotes the mixing time, i.e., the transition point between early reflections and late reverberation. As discussed in Sec.~\ref{subsec:representations}, small FDNs, like the one used in this analysis with $N=6$, tend to produce sparse early reflections. Compared to larger FDNs or real RIRs, substantial differences in the first few milliseconds of the response are inevitable, unless addressed using scattering elements \cite{schlecht2020scattering} or by implicitly modeling the target’s early reflections with FIR filters. We estimate $t_{\textrm{mix}}$ using the formula presented by Abel and Huang~\cite{abel2006simple}. Fig.~\ref{fig:loss_profile_II_edc} shows that the improvements in the resulting loss profiles, compared to Fig.~\ref{fig:loss_profile_I}, are minimal. Therefore, in the rest of the paper, we consider the entire response.


\subsection{Learning Filters from Noisy Room Impulse Responses}\label{subsec:learn_noisy_rirs}
The representations analyzed in Sections~\ref{subsec:representations} and~\ref{subsec:noise_loss_profile} confirm the assumption that the linear scale is more robust to noise than the logarithmic scale. Nonetheless, the relative MAE depicted in the first two rows of Table~\ref{tb:mae_perturbation}, as well as the loss profiles in Figures~\ref{fig:loss_profile_I} and~\ref{fig:loss_profile_II_edc}, show different levels of susceptibility to noise between EDC and MSS. In Sec.~\ref{sec:evaluation}, we focus on understanding how these results generalize to different targets, noise levels, and real data.

In particular, we demonstrate the effectiveness of the proposed optimization framework, in which gradient descent is used to minimize a dissimilarity metric evaluated between the noisy RIR and the sum of the FDN's IR $\hat{h}(t)$ and a noise sequence $w_2(t)$:
\begin{align}\label{eq:noisy_loss}
\mathcal{L}({h}+w_1, \hat{h}+w_2) = \mathcal{J}\Bigl(q\bigl({h}+w_1\bigr), q\bigl(\hat{h}+w_2\bigr)\Bigr)~.
\end{align}
Noise $w_2(t)$ is assumed to have the same statistical properties as $w_1(t)$ and to be mixed with $\hat{h}(t)$ to match the target SNR.
Furthermore, we will analyze how such noise-aware losses perform when combined with $\mathcal{L}_U$, which targets $\mat{U}$, a parameter that does impact the energy decay. 

\section{EVALUATION}\label{sec:evaluation}
Three experiments designed to evaluate the effects of parameter perturbation and noise modeling in gradient descent optimization are presented next. We focus on $\mathcal{L}_{\textrm{EDC,\,lin}}$, $\mathcal{L}_{\textrm{EDC,\,log}}$, and $\mathcal{L}_{\textrm{MSS}}$, computed over the full length of the signals, at resolutions specified in Sec.~\ref{subsec:noise_loss_profile}. In the first two experiments, the target response is always augmented with Gaussian noise at 10\,dB SNR, denoted as $w_1(t)$. This deliberately low SNR is chosen to test the loss functions under poor noise conditions and to expose potential failure of the training pipeline. Preliminary experiments with 10, 20, and 30\,dB SNR showed no qualitative differences in the loss profiles; we therefore report results for the worst-case setting. For all experiments, we use the \textit{FLAMO} library of differentiable LTI systems~\cite{santo2024flamo}. The code used to run all the experiments presented in this paper is available online\footnote{\url{https://github.com/gdalsanto/learn-recursive-filters}}.

\begin{table}[t!]
\centering
\caption{\textit{Relative MAE between the target parameters and the parameters corresponding
to the minima of the noise-aware loss functions under random perturbations.}}
\label{tb:mae_perturbation}

\begin{tabular}{ll
S[table-format=2.2]
S[table-format=2.2]
S[table-format=2.2]}
\toprule
\multicolumn{2}{c}{Parameters} &
\multicolumn{3}{c}{Relative MAE [\%] $\downarrow$} \\
\cmidrule(l){1-2} \cmidrule(l){3-5}
$\theta_{\mathrm{FI}}$ & $\theta_{\mat{\Gamma}}$ &
{$\mathcal{L}_{\mathrm{EDC,lin}}$} &
{$\mathcal{L}_{\mathrm{EDC,log}}$} &
{$\mathcal{L}_{\mathrm{MSS}}$} \\
\midrule

\multirow{2}{*}{---}
  & $\Tsixty^{\mathrm{dc}}$   & 4.99 & 7.62 & 27.24 \\
  & $f_\mathrm{c}$           & 5.40 & 5.84 & 32.75 \\
\addlinespace

\multirow{2}{*}{$\mat{b}$}
  & $\Tsixty^{\mathrm{dc}}$  & 51.83 & 30.69 & 27.96 \\
  & $f_\mathrm{c}$           & 39.83 & 34.38 & 25.36 \\
\addlinespace

\multirow{2}{*}{$\mat{c}$}
  & $\Tsixty^{\mathrm{dc}}$  & 72.87 & 34.77 & 73.25  \\
  & $f_\mathrm{c}$           & 73.16 & 68.79 & 75.33\\
\addlinespace

\multirow{2}{*}{$\mat{U}$}
  & $\Tsixty^{\mathrm{dc}}$  & 11.08 & 4.80 & 20.44 \\
  & $f_\mathrm{c}$           & 9.51 & 8.97 & 23.97 \\
\addlinespace

\multirow{2}{*}{$\mat{m}$}
  & $\Tsixty^{\mathrm{dc}}$  & 12.14 & 7.18 & 40.86 \\
  & $f_\mathrm{c}$           & 10.73 & 8.48 & 54.95 \\
\bottomrule
\end{tabular}
\end{table}

\subsection{Effect of Parameter Perturbation}\label{subsec:param_perturb}
We analyze the effects of perturbing the frequency-independent parameters $\theta_{\textrm{FI}}$ on the loss profile, following the method presented in Sec.~\ref{subsec:parameters_perturbation} and under the noise-aware condition at 10~dB SNR. The objective of this experiment is to analyze the sensitivity of the loss function to isolated parametric perturbations and to identify the parameters that cause the largest deviations from the minimum.

We use an FDN with $N=6$ delay lines of lengths $\mat{m} = [997, 1153, 1327, 1559, 1801, 2099]$ samples, yielding delays from 20.8~ms to 43.7~ms at $\fs=48$~kHz. We begin by sampling the parameters $\mat{b}$ and $\mat{c}$ as arrays of ones with alternating signs. We then scale them such that the energy at $\theta_{\mat{\Gamma}}=\theta_{\mat{\Gamma}}^*$ matches that of the target FDN. The orthogonal feedback matrix is generated by mapping an $N\times N$ matrix, drawn from $\mathcal{N}(0, 1)$, to the orthogonal space via skew-symmetrization and the matrix exponential \cite{lezcano2019cheap}. We use $M=196$~k frequency sampling points, which generate a response of 4\,s, ensuring minimal time aliasing. We compute the loss profile on 200 uniform steps within the parameter range $[\theta_{\mat{\Gamma}}', \theta_{\mat{\Gamma}}''] = [(0.25~\textrm{s}, 1.6~\textrm{kHz})),(3.75~\textrm{s}, 16~\textrm{kHz})]$, with targets $\Tsixty^{\textrm{dc *}} = 2~$s and $f_\textrm{c}^*=10$~kHz. At each step $\theta_{\mat{\Gamma}, j}$, we compute the loss over $K=50$ different instances of $\theta_{\textrm{FI}}$, sampled from a normal distribution with first-order statistics matched to those of their initial values. This sampling approach does not preserve energy matching at $\theta_{\mat{\Gamma}}=\theta_{\mat{\Gamma}}^*$, which allows us to evaluate the loss under more general conditions.

Table~\ref{tb:mae_perturbation} summarizes the relative mean absolute error (MAE) between the target parameters and the parameters corresponding to the minima of the loss functions. The loss profiles statistics are reported in Figs.~\ref{fig:perturb_bc} and \ref{fig:perturb_Um} of the \hyperref[appendix_perturbation]{Appendix}.
Perturbation of the input and output gains leads to the largest deviations in $\mathcal{L}_\textrm{EDC}$, most likely due to their strong influence on modal excitation and total energy. In particular, $\mat{c}$ drastically shifts the minima of all the losses toward $\theta_{\mat{\Gamma}}'$. The role of $\mat{c}$ is to combine the channels to form the mono output, and as such, it has a strong influence on the response energy. Feedback matrix $\mat{U}$ and delay-line lengths $\mat{m}$ introduce moderate deviations. For $\mathcal{L}_\textrm{MSS}$, perturbations of $\mat{U}$ seem to slightly improve the MAE. Overall, the results show clear differences in how $\mathcal{L}_\textrm{EDC}$ and $\mathcal{L}_\textrm{MSS}$ respond to perturbations of the frequency-independent parameters. $\mathcal{L}_\textrm{EDC,\,log}$ shows the highest robustness to these variations, whereas $\mathcal{L}_\textrm{MSS}$ is consistently sensitive to $\theta_{\textrm{FI}}$.

\subsection{Gradient-Descent Optimization}\label{subsec:grad_optim}
To evaluate the optimization of the attenuation filters and the performance of the noise-aware model, we perform gradient-descent optimization under four configurations: 
\begin{enumerate}
    \item $\mathcal{L}(h+w_1,\hat{h})$, optimize only $\theta_{\mat{\Gamma}}$ (keep $\theta_{\textrm{FI}}$ fixed).
    \item $\mathcal{L}(h+w_1,\hat{h})$, optimize both $\theta_{\mat{\Gamma}}$ and $\theta_{\textrm{FI}}$.
    \item $\mathcal{L}(h+w_1,\hat{h}+w_2)$, optimize only $\theta_{\mat{\Gamma}}$ (keep $\theta_{\textrm{FI}}$ fixed).
    \item $\mathcal{L}(h+w_1,\hat{h}+w_2)$, optimize both $\theta_{\mat{\Gamma}}$ and $\theta_{\textrm{FI}}$.
\end{enumerate}
The noise sequence added to the learnable FDN, $w_2(t)$, has the same total energy as $w_1(t)$. In all cases, we combine the objective with $\mathcal{L}_{U}$ to simulate a practical scenario in which the attenuation filters of an FDN are optimized alongside other criteria. 

We run $80$ independent optimization trials. Each trial uses a different target response, where the target attenuation parameters are drawn from a uniform distribution, $\Tsixty  \sim \mathcal{U}(1, 3.5)$~s and $f_\textrm{c} \sim \mathcal{U}(6, 12)$\,kHz. The target FDN is constructed as in Sec.~\ref{subsec:representations}. After sampling the target attenuation values, we initialize the learnable FDN as follows:
\begin{enumerate}
\item  Sample all frequency-independent parameters as described in Sec.~\ref{subsec:param_perturb}. Choose delay-line lengths (in samples) as pairwise co-prime integers corresponding to time delays in the range [15,\,45] ms. 
\item Compute the energy at the target attenuation setting $\theta^*_{\mat{\Gamma}}$.
\item  Scale the input and output gains $\mat{b}$ and $\mat{c}$ to match the energy of the target FDN's IR at $\theta^*_{\mat{\Gamma}}$
\item Randomly initialize $\theta_{\mat{\Gamma}}$.
\end{enumerate}
We optimize with Adam for up to 8000 iterations, divided into 40 epochs. We use early stopping with a patience of 10 epochs and a minimum improvement of $10^{-4}$ in the objective.  

Table~\ref{tb:mae_optimization} reports the results. The MAE for tests 3–4 is consistently lower, highlighting the benefit of modeling the noise. Overall, the EDC representation performs better, with the linear-scale variant showing particularly low MAE in tests 1–2, when noise is not modeled. In the noise-aware case, $\mathcal{L}_\textrm{MSS}$ and $\mathcal{L}_{\textrm{EDC, lin}}$ benefit more when the frequency-independent parameters are allowed to vary during the optimization. 

{\setlength{\tabcolsep}{3pt}
\begin{table}[t!]
\centering
\caption{\textit{Relative MAE between the target parameters and the parameters learned via gradient-descent optimization. Each test configuration, listed under Test, is defined in Sec.~\ref{subsec:grad_optim}.}}
\label{tb:mae_optimization}

\begin{tabular}{l llcccc}
\toprule
\multicolumn{3}{c}{Test} &
\multicolumn{4}{c}{Relative MAE [\%] $\downarrow$} \\
\cmidrule(l){1-3} \cmidrule(l){4-7}
\# & Model & Optim &
$\theta_{\mat{\Gamma}}$ &
$\mathcal{L}_{\mathrm{EDC,lin}}$ &
$\mathcal{L}_{\mathrm{EDC,log}}$ &
$\mathcal{L}_{\mathrm{MSS}}$ \\
\midrule

\multirow{2}{*}{1}
 & \multirow{2}{*}{$\hat{h}$}
 & \multirow{2}{*}{$\theta_{\mat{\Gamma}}$} & $\Tsixty^{\mathrm{dc}}$ & 5.15 & 120.59 & 105.04 \\
 &  &  & $f_\mathrm{c}$ & 15.87 & 164.43 & 160.0 \\
\addlinespace

\multirow{2}{*}{2}
 & \multirow{2}{*}{$\hat{h}$}
 & \multirow{2}{*}{$\theta_{\mat{\Gamma}}, \theta_{\textrm{FI}}$} & $\Tsixty^{\mathrm{dc}}$ & 12.20 & 104.43 & 124.14 \\
 &  &  & $f_\mathrm{c}$ & 14.38 & 162.98 & 164.02 \\
\addlinespace

\multirow{2}{*}{3}
 & \multirow{2}{*}{$\hat{h}+w_2$}
 & \multirow{2}{*}{$\theta_{\mat{\Gamma}}$} & $\Tsixty^{\mathrm{dc}}$ & 4.00 &  3.87 & 15.59 \\
 &  &  & $f_\mathrm{c}$ & 3.45 &  4.18 & 28.54 \\
\addlinespace

\multirow{2}{*}{4}
 & \multirow{2}{*}{$\hat{h}+w_2$}
 & \multirow{2}{*}{$\theta_{\mat{\Gamma}}, \theta_{\textrm{FI}}$} & $\Tsixty^{\mathrm{dc}}$ & 2.88 & 5.14 & 5.47 \\
 &  &  & $f_\mathrm{c}$ & 1.85 & 4.62 & 4.90 \\
\bottomrule
\end{tabular}
\end{table}}

\subsection{Fitting Real Data}\label{sec:fitting_real}
Finally, we validate our approach on a subset of measured RIRs and background noise recordings. The differentiable FDN structure is extended by introducing a tone-correction filter outside the feedback loop. Specifically, we use a GEQ composed of low- and high-shelving filters with crossover frequencies of 44~Hz and 22.6~kHz, respectively, and one-octave bands with center frequencies from 62.5~Hz to 16~kHz. The filter is initialized with a flat 0~dB response, and the band command gains are treated as optimizable parameters to match the overall spectral energy of the measured RIRs.

To simulate background noise, we use three recordings from the TAU dataset~\cite{politis_2022_6408611}. These recordings were captured in an underground shelter, a small classroom, and a large classroom, all dominated by ventilation noise. From each noise recording, we sample a segment free of transient events. We then add the noise segments to the target RIR to obtain either 10~dB or 20~dB SNR. In addition, we run experiments with one instance of white Gaussian noise at the same SNR levels.

We use seven mono RIRs selected from two datasets \cite{lachenmayr2023chamber, sergio_ok5}. The subset includes two hallways, one concert hall, a chamber music hall, a lecture hall, a miscellaneous room, and a bathroom. Across eight frequency bands from 125~Hz to 16~kHz, the mean $\Tsixty$ ranges from 0.54~s to 2.19~s. In total, for each SNR level and loss function, we run optimization on 28 noisy RIRs. 

We model the noise using the following procedure:
\begin{itemize}
    \item Detect the intersection time between the RIR decay and the background noise floor using PyRato's implementation of the Lundeby method~\footnote{\url{https://pyrato.readthedocs.io/en/stable/pyrato.html}}.
    \item Extract a noise segment from the tail of the RIR starting at the intersection time.
    \item Extend the extracted noise segment to cover the full RIR duration while preserving its spectral characteristics via phase randomization~\cite{valimaki2018creating}.
\end{itemize}

During optimization, all parameters except $\mat{m}$ were updated via gradient descent. We used the same training configuration as in Sec.~\ref{subsec:grad_optim}. We additionally applied a multiplicative factor of 3 to $\mathcal{L}_{{U}}$ to compensate for the larger discrepancy between the reference and estimated RIRs, particularly at initialization. Before optimization, $\mat{c}$ and $\mat{b}$ are scaled to match the energy of the target RIR, obtained by subtracting the energy of the extracted noise from that of the noisy target RIR. The FDN, however, is still free to change its overall energy during training.

\begin{figure}
    \centering
    \includegraphics[width=0.5\linewidth]{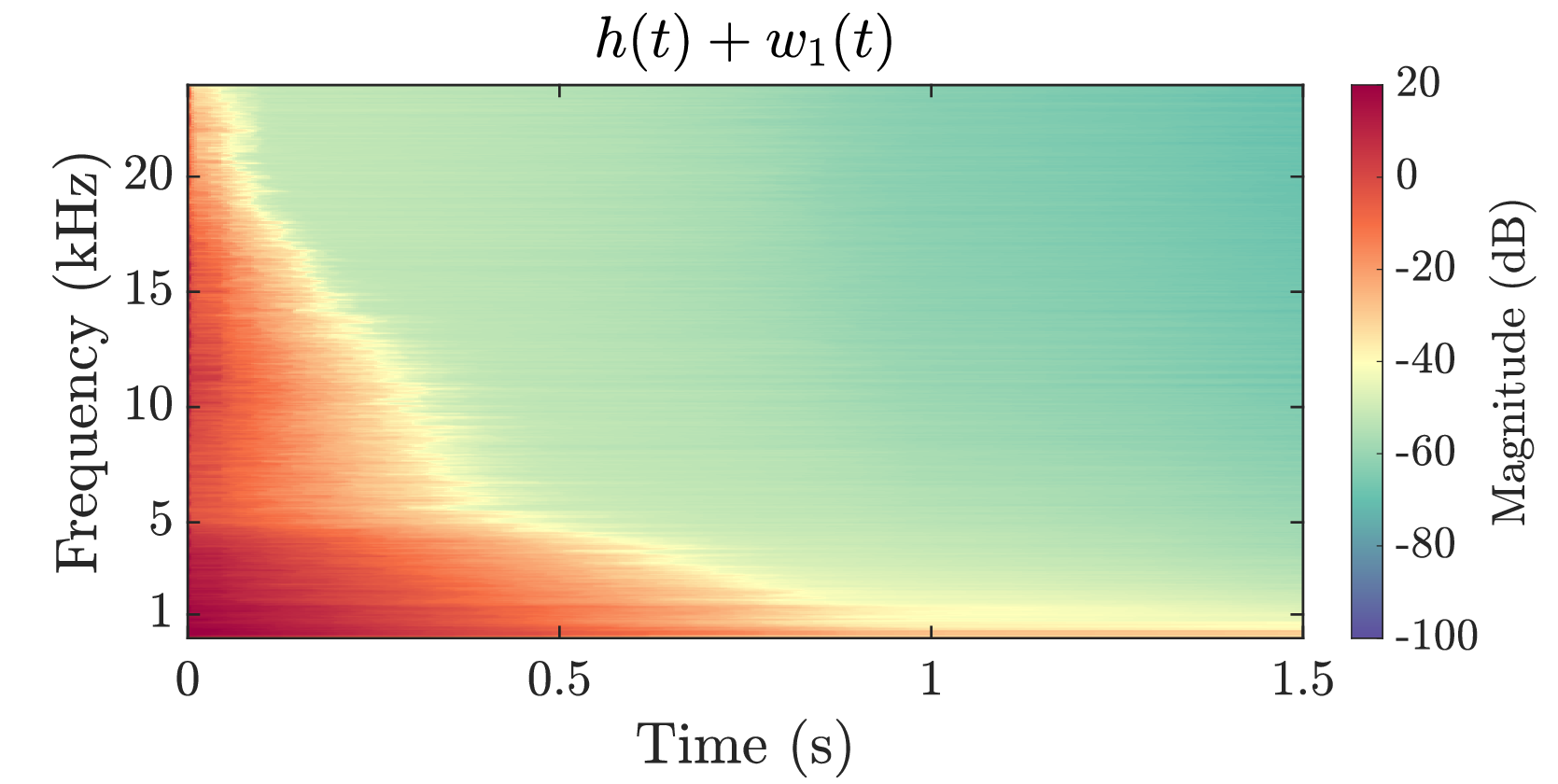}
    \caption{2D projection of the EDR of the target clean RIR measured in a chamber music hall.}
    \label{fig:ref_edr}
\end{figure}

\begin{figure}[h]
    \centering
    \begin{subfigure}{\linewidth}
        \centering

        \includegraphics[width=0.65\linewidth]{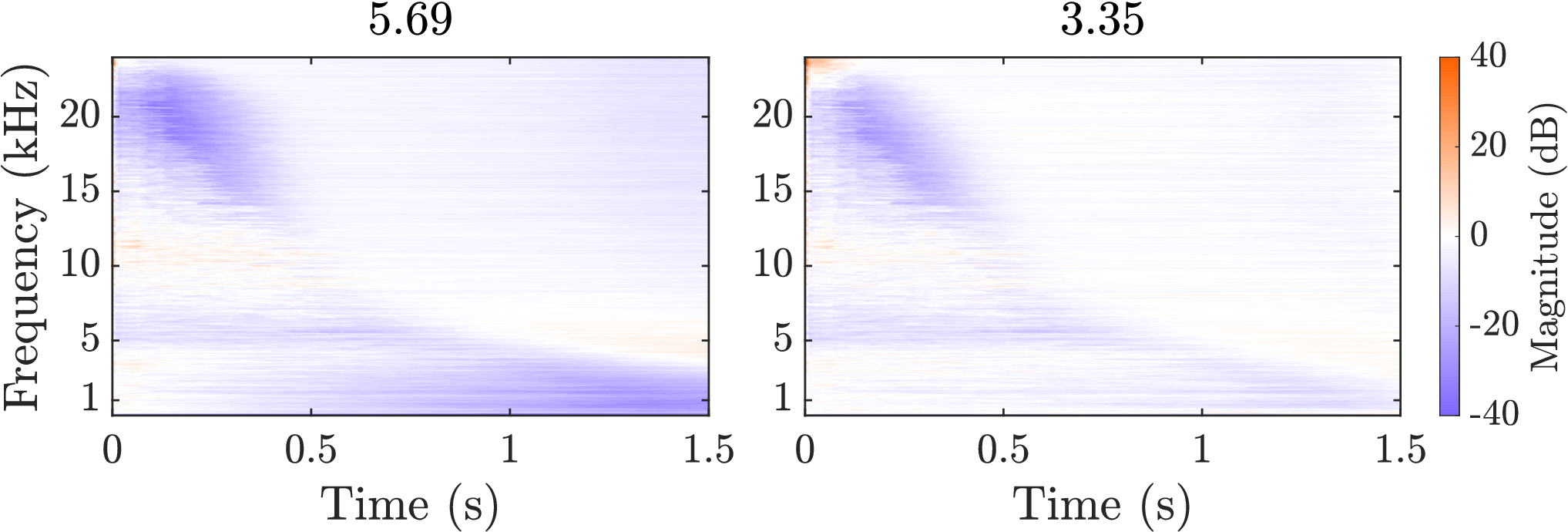}
        \caption{$\mathcal{L}_{\mathrm{EDC,lin}}$}
        \label{fig:perturb_U}
        \vspace{0.1cm}
    \end{subfigure}
    
    \begin{subfigure}{\linewidth}
        \centering

        \includegraphics[width=0.65\linewidth]{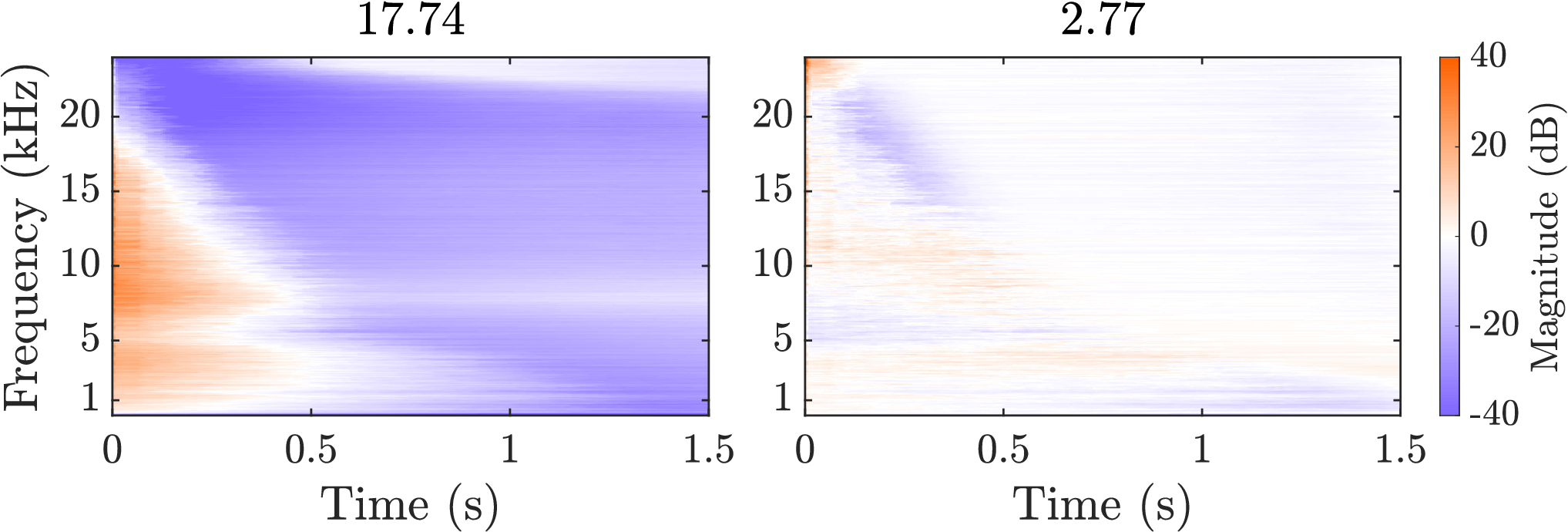}
        \caption{$\mathcal{L}_{\mathrm{EDC,log}}$}
        \label{fig:perturb_U}
        \vspace{0.1cm}
    \end{subfigure}

    \begin{subfigure}{\linewidth}
        \centering
        \includegraphics[width=0.65\linewidth]{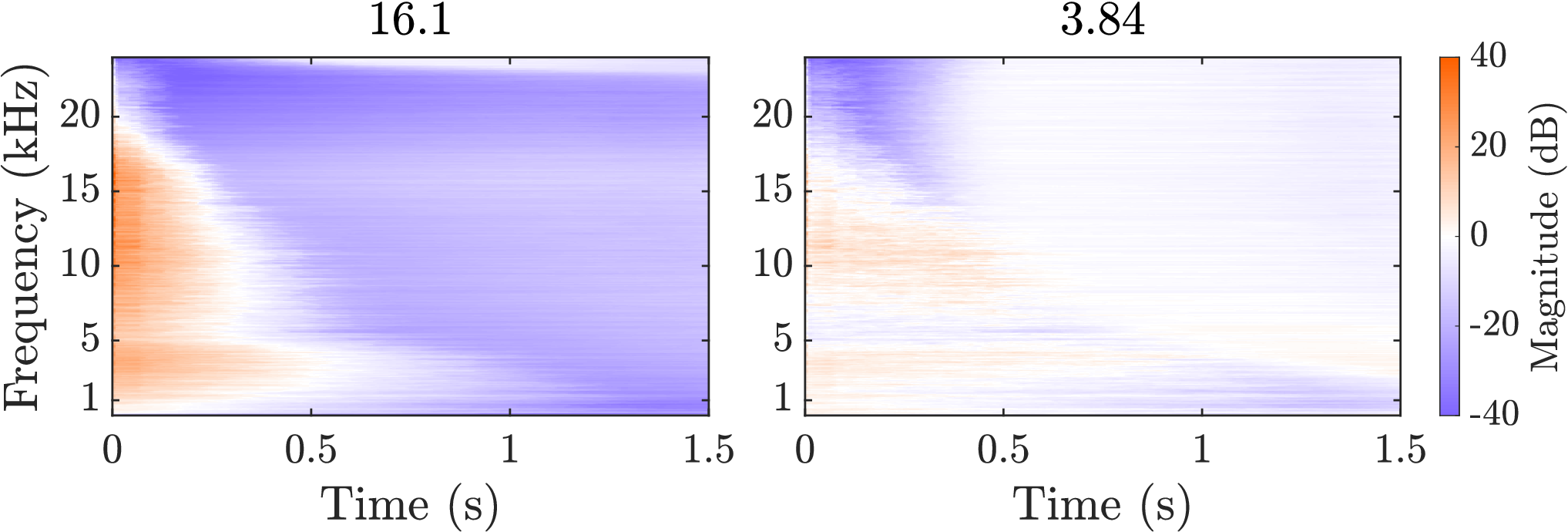}
        \caption{$\mathcal{L}_{\mathrm{MSS}}$}
        \label{fig:perturb_m}
    \end{subfigure}
    \caption{Difference between the EDR of the clean target RIR in Fig.~\ref{fig:ref_edr} and estimated RIR using noise-agnostic (left) and noise-aware (right) distances. Subplots are grouped by loss function. Numbers above the plots indicate the MAE.}
    \label{fig:edr_diff}
\end{figure}

We report in Fig.~\ref{fig:ref_edr} the 2D projection of the energy decay relief (EDR) \cite{jot1992analysis} of one measured RIR. The EDR is based on Schroeder backward integration of the STFT rather than the IR and was used in \cite{lee2022differentiable} to quantify the performance of differentiable reverberators. 
As an example, the EDR difference in dB between one measured RIR and the optimized FDN's IR is shown in Fig.~\ref{fig:edr_diff}. All EDRs have been computed from the noise-free responses to compare better the effect of the proposed method on the learning of the FDN parameters. Overall, all losses show improvements in the noise-aware case, with the remaining errors most likely due to the inability of the low-pass attenuation filter to model finer nuances in the decay rate. 
Audio examples of the target and estimated RIRs are available online\footnote{\url{http://research.spa.aalto.fi/publications/papers/jaes-learn-noisy-rirs/}}. 

The MAE between the EDR in dB of the target RIRs and that of the estimated RIRs is reported in Table~\ref{tb:fitting_real}, where the results are organized by SNR level and noise type. The MAE is computed across time and frequency bins and then averaged over all RIR-noise pairs. Results for white Gaussian noise and TAU noise are reported separately. While the noise-agnostic method shows improvement at the higher SNR level, the noise-aware losses present similar values across the two SNRs. This result suggests that our methodology is more robust to varying noise conditions. All loss functions benefited from the proposed approach, with $\mathcal{L}_{\textrm{EDC, log}}$ exhibiting the greatest improvement overall. White Gaussian noise produces higher errors in all tested conditions. 

\setlength{\tabcolsep}{3pt}
\begin{table*}[t!]
\centering

\caption{\textit{Mean MAE between the EDR (in dB) of the target RIR and that of the FDN IR learned via gradient-descent optimization (Sec.~\ref{sec:fitting_real}). Numbers in brackets indicate the standard deviation. Results for white Gaussian noise and TAU noise are reported separately.}}
\label{tb:fitting_real}

\begin{tabular}{llcccccccccccc}
\toprule
\multicolumn{2}{c}{} &
\multicolumn{12}{c}{MAE [dB] $\downarrow$} \\
\cmidrule(l){3-14}
\multicolumn{2}{c}{Test} &
\multicolumn{6}{c}{Gaussian} & \multicolumn{6}{c}{TAU} \\
\cmidrule(lr){1-2}\cmidrule(lr){3-8} \cmidrule(lr){9-14}

SNR & Model &
\multicolumn{2}{c}{\;\;$\mathcal{L}_{\mathrm{EDC,lin}}\;\;$ }&
\multicolumn{2}{c}{\;\;$\mathcal{L}_{\mathrm{EDC,log}}\;\;$ }&
\multicolumn{2}{c}{\;\;$\mathcal{L}_{\mathrm{MSS}}\;\;$}&
\multicolumn{2}{c}{\;\;$\mathcal{L}_{\mathrm{EDC,lin}}\;\;$ }&
\multicolumn{2}{c}{\;\;$\mathcal{L}_{\mathrm{EDC,log}}\;\;$ }&
\multicolumn{2}{c}{\;\;$\mathcal{L}_{\mathrm{MSS}}\;\;$} \\
\cmidrule(l){1-2}\cmidrule(l){3-8} \cmidrule(l){9-14}
\multirow{2}{*}{10 dB}

& $\hat{h}$ 

& 9.80 & (3.40)
& 53.28 & (3.65)
& 39.50 & (7.52) 
& 6.32 & (2.82)
& 11.29 & (6.67)
& 9.48 & (4.70) \\

& $\hat{h}+w_2$
& 7.02 & (2.38)
& 6.64 & (2.31)
& 6.03 & (2.22) 
& 5.71 & (2.85)
& 4.89 & (2.60)
& 5.10 & (2.31) \\
\cmidrule(l){1-2} \cmidrule(l){3-8} \cmidrule(l){9-14}
\multirow{2}{*}{20 dB} &

$\hat{h}$
& 6.94 & (2.56)
& 47.22 & (5.08)
& 32.63 & (4.62)
& 5.83 & (2.90)
& 12.28 & (1.38)
& 10.25 & (1.85)\\

& $\hat{h}+w_2$
& 6.80 & (2.63)
& 6.04 & (1.97)
& 6.04 & (2.08)
& 5.77 & (2.86)
& 4.80 & (2.63)
& 5.12 & (2.34) \\
\bottomrule
\end{tabular}
\end{table*}

\section{DISCUSSION}\label{sec:discussion}

The results reported above show that explicitly modeling the background noise of RIR improves the accuracy of the estimated attenuation filter in FDNs. Moreover, in our experiments, noise-aware configurations (tests 3 and 4) converged faster than the noise-agnostic ones (tests 1 and 2): on average, tests 3 and 4 required respectively 17.51 and 30.94 epochs to reach the early-stopping criterion, compared with 32.38 and 35.86 epochs for tests 1 and 2. We interpret this as noise acting to regularize the optimization landscape: modeling the noise reduces systematic mismatch between model and measurements, which yields faster, more stable parameter updates. As expected, when both parameter sets are tuned jointly, the increased dimensionality of the optimization problem causes the loss to require more iterations to settle. However, for $\mathcal{L}_{\textrm{EDC, lin}}$ and $\mathcal{L}_{\textrm{MSS}}$, changes in $\theta_{\textrm{FI}}$ lead to improved performance. When combined with Table~\ref{tb:mae_perturbation}, the results in Table~\ref{tb:mae_optimization} could indicate that a more robust loss function, such as $\mathcal{L}_{\textrm{EDC, log}}$, can be detrimental when used in combination with losses that target $\theta_{\textrm{FI}}$. Intuitively, in the noise-aware case, a more static loss profile could cause the optimization to overfit to the noise. 
Nonetheless, more experiments varying $\theta_{\textrm{FI}}$ are necessary to confirm this intuition.

Since measured RIRs typically exhibit lowpass behavior, as shown in Fig.~\ref{fig:ref_edr}, the background noise can cover large portions of the STFT (see synthesized example in Fig.~\ref{fig:stft_plots}), increasing the density of errors in frequency–time bins that are less perceptually relevant.  Allowing frequency-independent parameters to vary during optimization (Table~\ref{tb:fitting_real} and test 4 in Table~\ref{tb:mae_optimization}) mitigates this effect by giving the model additional freedom to balance the spectral energy.

The EDC-based representation benefits strongly from incorporating noise into the model: in the presence of Gaussian noise, the noise-aware tests show overall improvements in MAE with small differences between linear and logarithmic scales. Conversely, in the noise-agnostic case, the linear scale performed better (see Tables~\ref{tb:mae_optimization} and ~\ref{tb:fitting_real}), a result due to the logarithmic scale disproportionately amplifying differences near the noise floor. Overall, the linear EDC down-weights differences in low-amplitude regions (such as the noise floor), which makes it more robust when background noise is present but might prevent it to fit finer nuances when fitting on real data (Table~\ref{tb:fitting_real}). The results in Table~\ref{tb:fitting_real} also show that colored noise, such as that from the TAU dataset, presents an easier optimization scenario.

In this study, we used a low-order low-pass shelving filter, which yields smooth transitions between octave bands and motivates the use of one-octave resolution in the EDC. For more complex filter designs, such as GEQ and PEQ, a higher frequency resolution in the loss function is advisable. For even higher frequency resolutions, the EDR can be used. 
We expect the results to be similar to those obtained with EDC-based losses, given the commutative nature of the operations. 

In these experiments, we deliberately added the noise $w_2(t)$ to the system's IR at a predefined SNR level, or estimated the noise sequence from the tail of the measured RIR. Alternatively, a noise sequence with learnable spectral energy could be incorporated into the reverberation model. A detailed study of the training framework required for such an approach is left for future work.

Taken together, these findings indicate that: including a background-noise model is important both for accuracy and for faster convergence; in $\mathcal{L}_{\textrm{EDC}}$ the choice of the scale (linear vs. logarithmic) affects robustness to the noise floor; $\mathcal{L}_{\textrm{MSS}}$ yields competitive results only in the noise-aware setting and when additional degrees of freedom are permitted in the model.

\section{CONCLUSION}\label{sec:conclusion}

This study investigates gradient-based tuning of attenuation filters in FDNs from noisy RIR targets, focusing on two parameters of a low-order shelving filter: $\Tsixty$ at dc and the crossover frequency. We compared EDC- and MSS-based losses under background noise conditions and under perturbations of frequency-independent FDN parameters.

Evaluation of loss profiles reveals that, when optimization is performed against noisy targets without explicitly modeling the noise, the minima of both $\mathcal{L}_{\textrm{EDC}}$ and $\mathcal{L}_{\textrm{MSS}}$ shift away from the ground-truth attenuation parameters. Introducing a noise-aware objective, by adding a noise term to the model output to match the target SNR, substantially reduces this mismatch and improves optimization reliability across synthetic and real data. Moreover, while isolated perturbations in frequency-independent parameters increase estimation error, results from joint optimization indicate that allowing these parameters to vary can improve accuracy for $\mathcal{L}_{\textrm{EDC, lin}}$ and $\mathcal{L}_{\textrm{MSS}}$. 

These findings can inform the design of more accurate and reproducible FDN optimization pipelines under noisy conditions. Extending these conclusions beyond the low-order filter setting considered here, and jointly learning the SNR from noise data via gradient descent, remain directions for future work.

\section{ACKNOWLEDGMENT}
The Aalto University School of Electrical Engineering funded the work of the first author. This work was supported by the HUCE infrastructure of the Aalto University School of Electrical Engineering. GDS and SJS were supported through the joint German Academic Exchange Service (DAAD) and Research Council of Finland (RCF) Project (57763119) ``Immersive Augmented Acoustics (IAA).'' The computational resources were provided by the Aalto Science-IT project.
\bibliographystyle{alpha}
\bibliography{sample}
\break

\appendix

\section*{APPENDIX}
\subsection{Perturbation Analysis}\label{appendix_perturbation}
\begin{figure*}[!h]
    \centering

    \begin{subfigure}{\linewidth}
        \centering
        \includegraphics[width=0.75\linewidth]{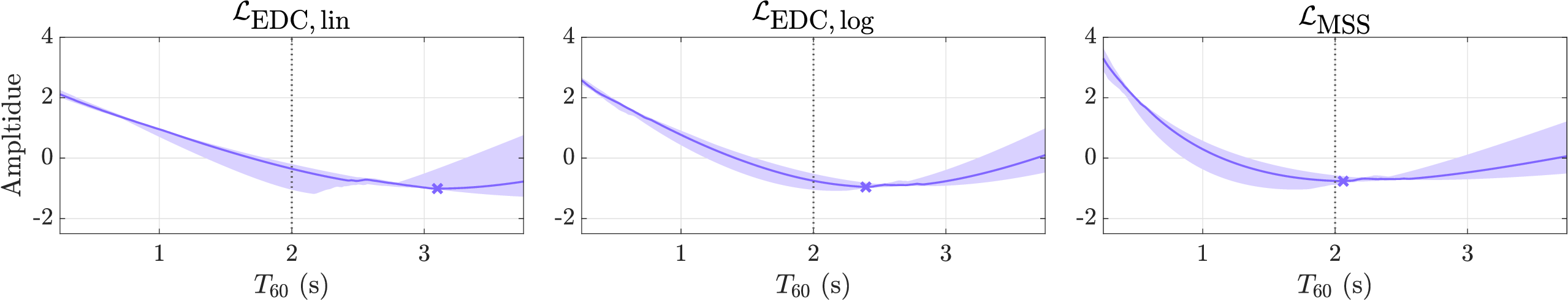} 
        \includegraphics[trim={0 0 0 0}, width=0.75\linewidth]{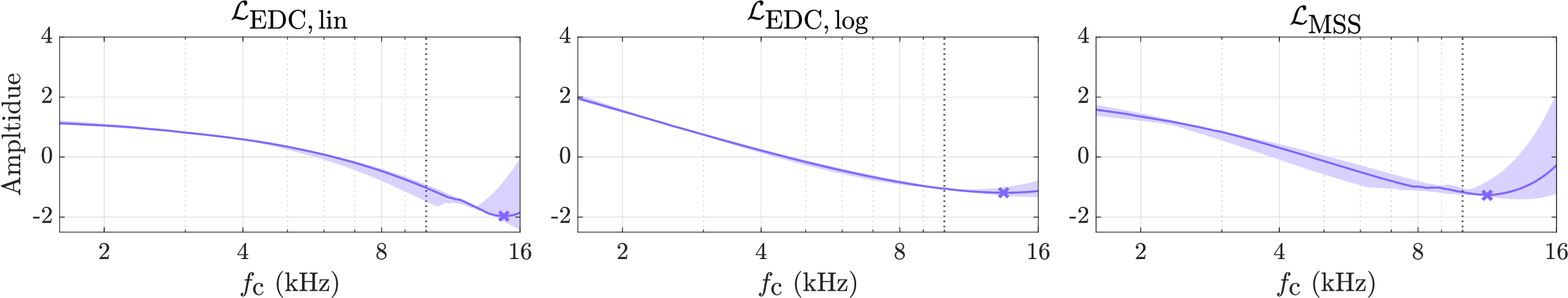}
        \caption{Input gains $\mat{b}$}
        \vspace{0.1cm}
        \label{fig:perturb_b}
    \end{subfigure}

    \begin{subfigure}{\linewidth}
        \centering
        \includegraphics[width=0.75\linewidth]{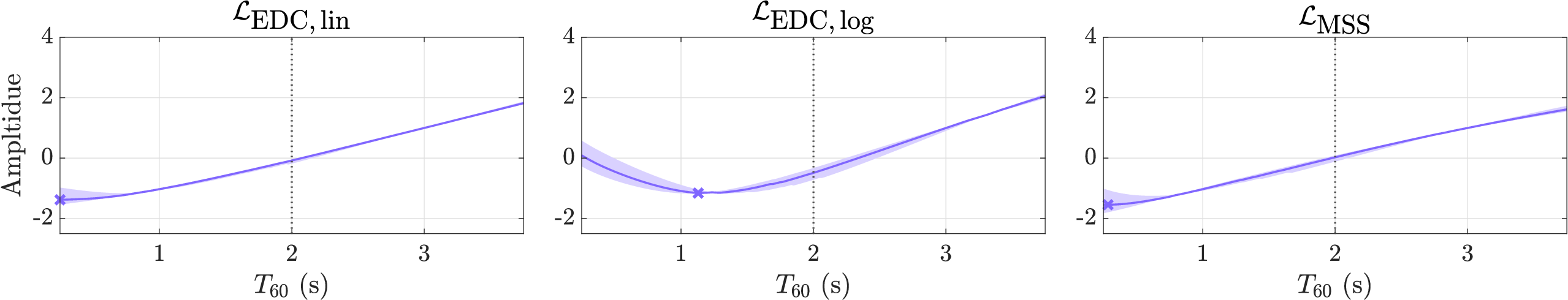}
        \includegraphics[width=0.75\linewidth]{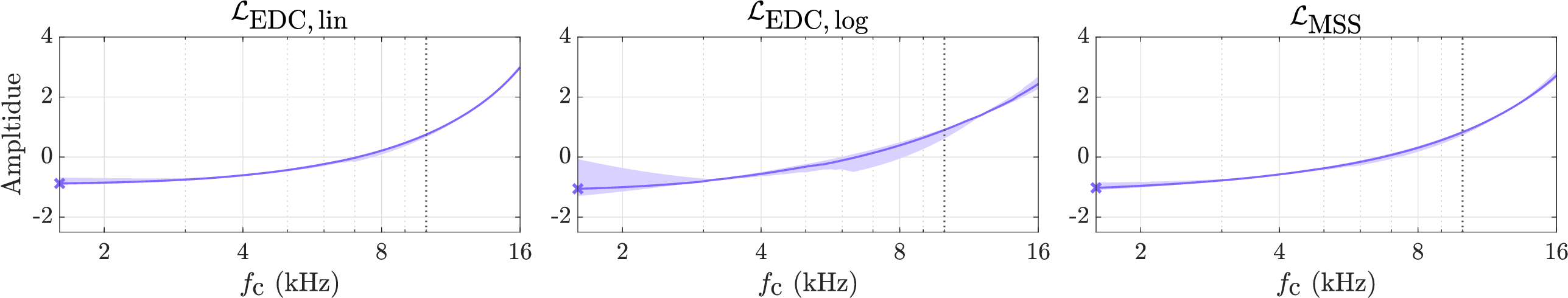}
        \caption{Output gains $\mat{c}$}
        \label{fig:perturb_c}
    \end{subfigure}

    \caption{Loss profile for 200 steps between two states of $\theta_{\mat{\Gamma}} = (\Tsixty^{\textrm{dc}}, f_\textrm{c})$. The solid line represents the median over $K=50$ instances of the input gain $\mat{b}$ (a) and output gain $\mat{c}$ (b). The shaded region indicates the range between the first and third quartiles. The targets $\Tsixty^{\textrm{dc}~*} = 2$~s and $f_\textrm{c}^*=10$~kHz are indicated with a vertical dotted line. The curve minimum is indicated with $\times$.}
    \label{fig:perturb_bc}
\end{figure*}

\begin{figure*}[h]
    \centering
    
    \begin{subfigure}{\linewidth}
        \centering

        \includegraphics[width=0.75\linewidth]{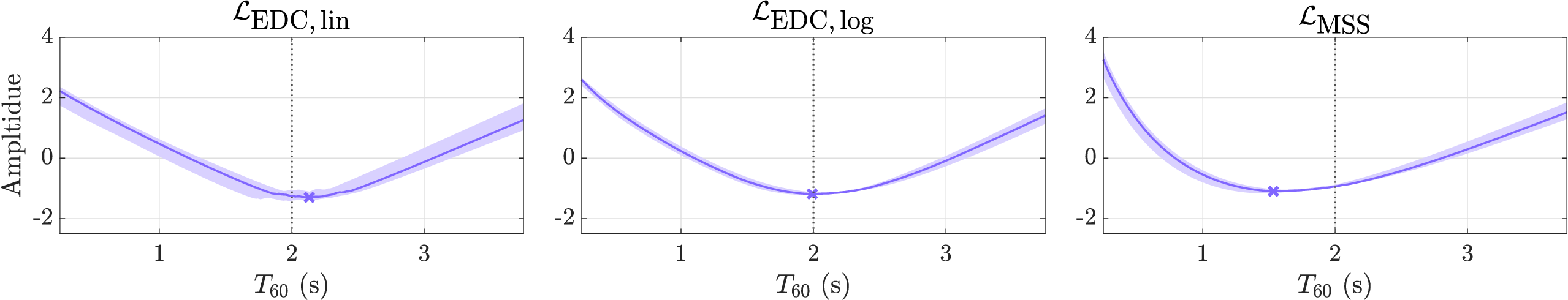}
        \includegraphics[width=0.75\linewidth]{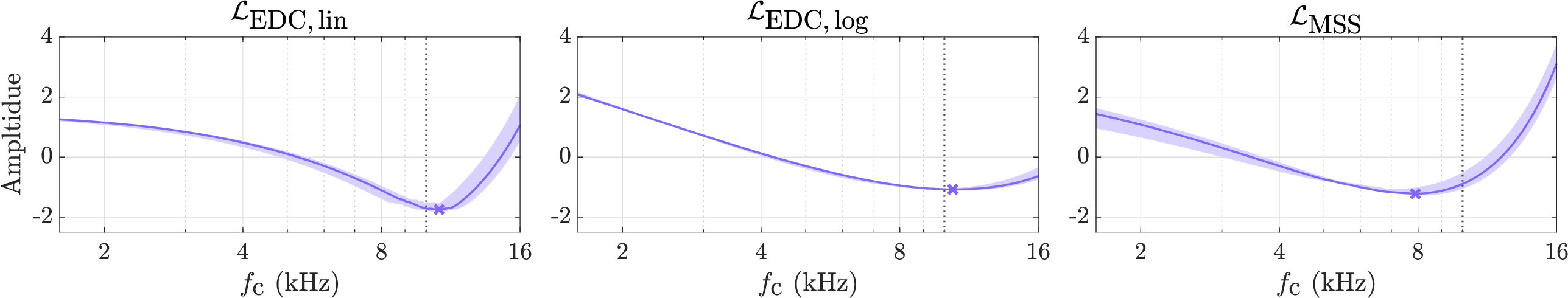}
        \caption{Feedback matrix $\mat{U}$}
        \label{fig:perturb_U}
        \vspace{0.1cm}
    \end{subfigure}

    \begin{subfigure}{\linewidth}
        \centering
        \includegraphics[width=0.75\linewidth]{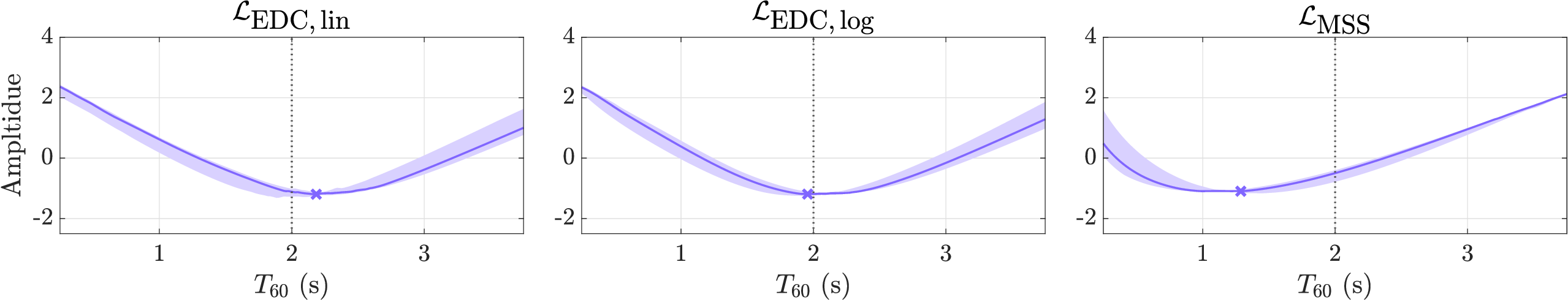}
        \includegraphics[width=0.75\linewidth]{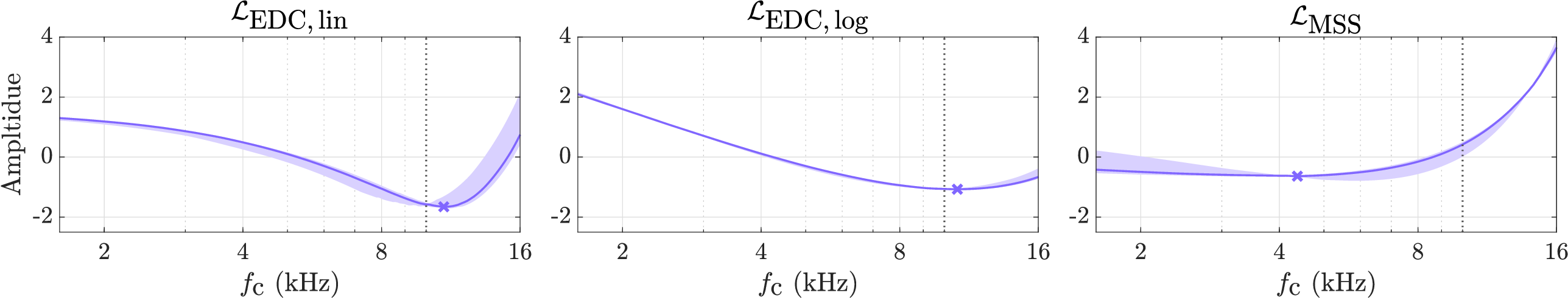}
        \caption{Delay lines $\mat{m}$}
        \label{fig:perturb_m}
    \end{subfigure}
    \caption{Loss profile for 200 steps between two states of $\theta_{\mat{\Gamma}} = (\Tsixty^{\textrm{dc}}, f_\textrm{c})$.
    The solid line represents the median over $K=50$ instances of the feedback matrix $\mat{U}$ (a) and delay lines $\mat{m}$ (b). The shaded region indicates the range between the first and third quartiles. The targets $\Tsixty^{\textrm{dc}~*} = 2$~s and $f_\textrm{c}^*=10$~kHz are indicated with a vertical dotted line. The curve minimum is indicated with $\times$.}
    \label{fig:perturb_Um}
\end{figure*}

Figures~\ref{fig:perturb_bc} and~\ref{fig:perturb_Um} illustrate the loss profiles relative to the results reported in Table~\ref{tb:mae_perturbation}. Figure~\ref{fig:perturb_c} demonstrates that the minimum of all loss functions is highly affected by $\mat{c}$. One hypothesis is that the random perturbation applied to $\mat{c}$ does not preserve its norm, thus changing the response energy. For the other parameters, the interquartile range decreases near the minimum, indicating the robustness of the losses under parameter perturbations. One exception is $\mathcal{L}_{\textrm{EDC, lin}}$ over perturbations of $\mat{b}$, as shown in Fig.~\ref{fig:perturb_b}.  Because the perturbations were constructed to preserve the mean of the parameter values, this result suggests that the losses are invariant to different linear combinations of the output channels.

\subsection{Loss Surface Analysis}\label{appendix_surface}
\begin{figure*}[h]
    \centering
    
    \begin{subfigure}{\linewidth}
        \centering
        \includegraphics[width=0.75\linewidth]{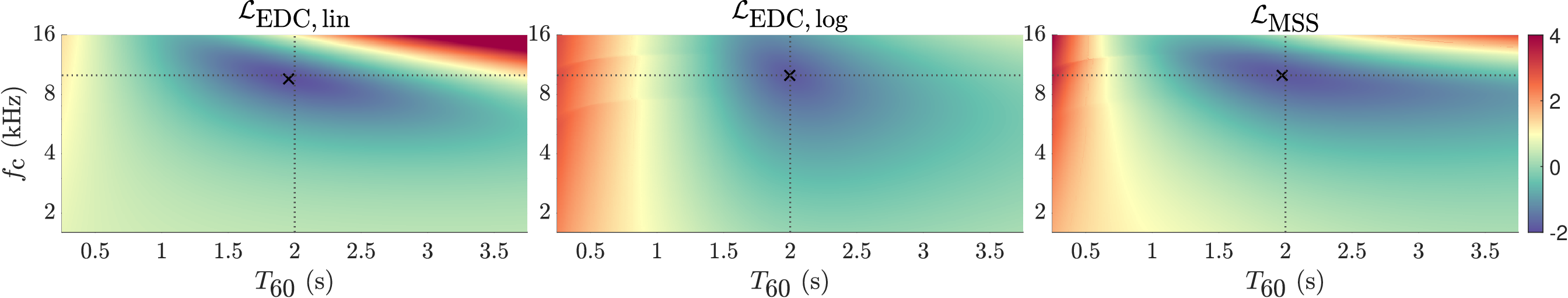}
        \caption{$\mathcal{L}(h, \hat{h}))$}
        \label{fig:no_noise}
        \vspace{0.1cm}
    \end{subfigure}

    \begin{subfigure}{\linewidth}
        \centering
        \includegraphics[width=0.75\linewidth]{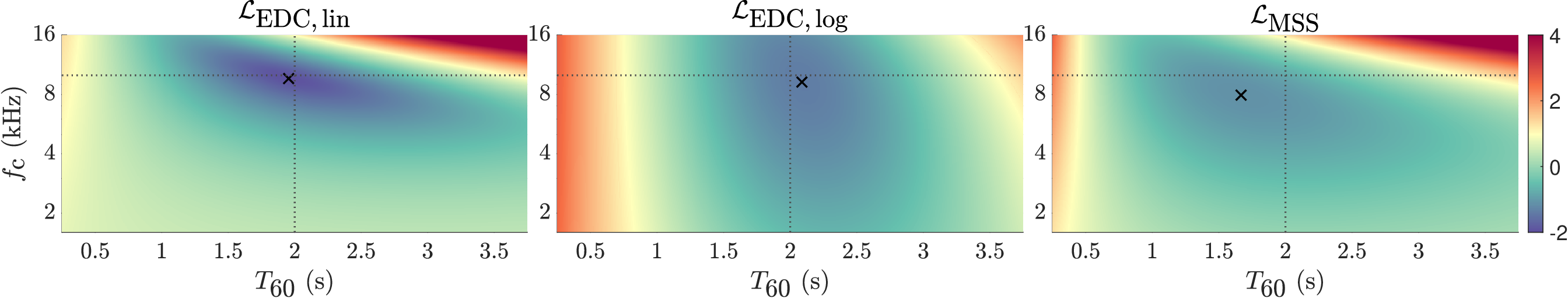}
        \caption{$\mathcal{L}(h + w_1, \hat{h}+w_2)$}
        \label{fig:noise_aware}
    \end{subfigure}
    
    \begin{subfigure}{\linewidth}
        \centering
        \includegraphics[width=0.75\linewidth]{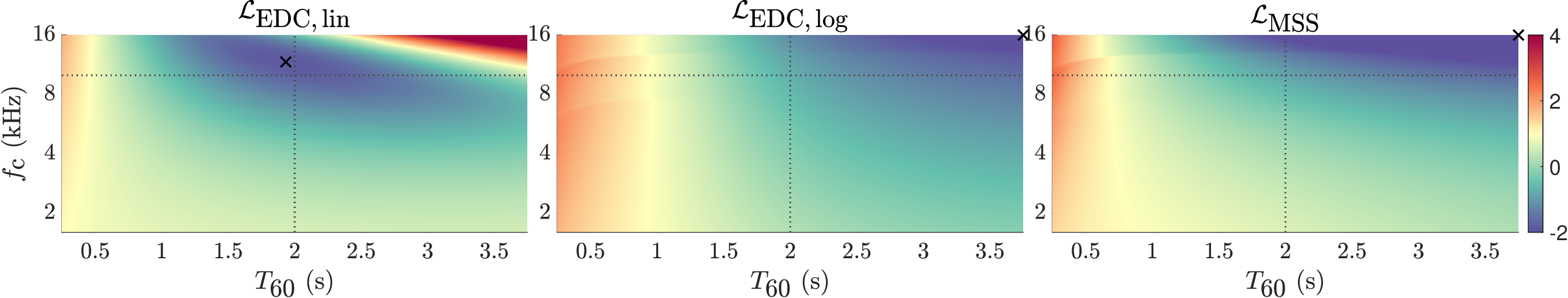}
        \caption{$\mathcal{L}(h + w_1, \hat{h})$}
        \label{fig:noise_agnostic}
    \end{subfigure}

    \caption{Loss surface plots for 500 steps between two states of $\theta_{\mat{\Gamma}} = (\Tsixty^{\textrm{dc}}, f_\textrm{c})$. The targets $\Tsixty^{\textrm{dc}~*} = 2$~s and $f_\textrm{c}^*=10$~kHz are indicated with dotted lines. Figures are divided by noise condition. The surface minimum is indicated with $\times$. For ease of comparison, all loss surfaces were standardized to zero mean and unitary standard deviation.}
    \label{fig:loss_surface}
\end{figure*}

The loss profile analysis in Sec.~\ref{subsec:noise_loss_profile} demonstrates convexity and smoothness of the loss only along a fixed line in one of the two parameters in $\theta_{\mat{\Gamma}}$. Before running the joint optimization of $T_{60}$ and $f_\textrm{c}$, it should be verified that the full loss surface exhibits a single minimum and that this minimum lies along both lines defined by the target parameter values. Figure~\ref{fig:loss_surface} shows the two-dimensional counterparts of the loss profiles in Fig.~\ref{fig:loss_profile_I}, retaining the same background noise configurations but using different instances of the FDNs and noise sequences. All loss surfaces were individually standardized to zero mean and unitary standard deviation. In the noiseless condition (Fig.~\ref{fig:no_noise}), this minimum lies very close to the target values. Similarly to the one-dimensional case, when noise is present only in the target signal (Fig.~\ref{fig:noise_agnostic}), the minimum deviates considerably, particularly for $\mathcal{L}_\textrm{EDC, log}$ and $\mathcal{L}_\textrm{MSS}$. This deviation is reduced in the \textit{noise-aware} condition (Fig.~\ref{fig:noise_aware}).

\end{document}